\documentclass[11pt,a4paper]{article}
\usepackage{jcappub}

\usepackage{color}
\usepackage{amsmath}
\usepackage{epsfig}
\usepackage{array}

\title{Keeping it real: revisiting a real-space approach to running ensembles of cosmological N-body simulations
}
        
\author[a,b]{ Chris Orban}

\affiliation[a]{Center for Cosmology and Astro-Particle Physics, The Ohio State University, 191 W Woodruff Ave, Columbus, OH 43210 } 
\affiliation[b]{Department of Physics, The Ohio State University, 191 W Woodruff Ave, Columbus, OH 43210 } 

\emailAdd{orban@physics.osu.edu}


\arxivnumber{1201.2082}

\abstract{
In setting up initial conditions for ensembles of cosmological N-body simulations there are,
fundamentally, two choices: either maximizing the correspondence of the initial
density field to the assumed fourier-space clustering or, instead,
matching to real-space statistics and allowing the DC mode (i.e. overdensity)
to vary from box to box as it would in the real universe. As a stringent test of both approaches,
I perform ensembles of simulations using power law and a ``powerlaw times a bump'' model 
inspired by baryon acoustic oscillations (BAO), exploiting the self-similarity of these initial 
conditions to quantify the accuracy of the matter-matter two-point
correlation results. The real-space method, which was originally proposed by 
Pen 1997 \cite{Pen1997} and implemented by Sirko 2005 \cite{Sirko2005},
 performed well in producing the expected self-similar behavior and corroborated the non-linear evolution 
of the BAO feature observed in conventional simulations, even in the 
strongly-clustered regime ($\sigma_8 \gtrsim 1$). In revisiting the real-space
method championed by \cite{Sirko2005}, it was also noticed that this earlier study overlooked an 
important integral constraint correction to the correlation function 
{\it in results from the conventional approach} that can be important in $\Lambda$CDM 
simulations with $L_{\rm box} \lesssim 1 \, h^{-1}$Gpc and on scales $r \gtrsim L_{\rm box} / 10$. 
Rectifying this issue shows that the fourier space and real space methods 
are about equally accurate and efficient for modeling the evolution and growth of 
the correlation function, contrary to previous claims.
An appendix provides a useful independent-of-epoch analytic formula 
for estimating the importance of the integral constraint bias on correlation function measurements in 
$\Lambda$CDM simulations.}

\begin{document}
\maketitle

\section{Introduction}
  \label{sec:intro}

Next-generation astronomical surveys will demand increasingly precise 
predictions from theory in order to properly interpret observations and
constrain the nature of dark energy. As emphasized by \cite{Annis_etal05,Smith_etal2012},
this will be a challenging task:  inaccuracies in the predictions of
halo abundance and halo bias, for example, can affect cosmological 
inferences \citep{Heidi_etal2010,Reed_etal2013}, and measurements of the baryon acoustic 
oscillations (BAO) clustering feature will soon reach the stage where 
theoretical estimates of the shift of this feature from non-linear dynamics 
become important \citep{Seo_etal2009}.  
Although  current state-of-the-art 
cosmological N-body simulations, given a specific set of cosmological parameters, 
are in many ways well-equipped to deliver highly precise predictions of the dark matter 
two-point correlation function and power spectrum for a relatively wide range of scales
\citep{Heitmann_etal2010}, the difficult-to-estimate covariances of these 
statistics are also crucial for placing constraints on cosmological parameters 
\citep{Habib_etal2007,Takahashi_etal2009}. While much creativity has gone
into methods and algorithms that ultimately save substantial computer time in delivering
these predictions \cite{Schneider_etal2008,Schneider_etal2011,Angulo_White2010,Tassev_Zaldarriaga2012}
the cosmological $N$-body simulations that these methods draw upon, 
with very few exceptions, are conducted \emph{without} allowing the overdensity
in each box to vary as it would if boxes of a cosmologically-relevant size were randomly 
positioned in the universe. The goal of this paper is twofold: (1) to assess the 
ramifications of this choice and in doing so explore the predictions of the conventional (or ``standard'')
method, (2) to explore the predictions of a method that
does allow the overdensity to vary from box-to-box or otherwise to 
document -- for lack of existing references -- why the field has largely
abandoned this approach.

Since other authors have adequately described the conventional
method \cite{Efstathiou_etal1985,Peacock1999,Martel2005}, which seeks to maximize
the correspondence between the assumed initial fourier space clustering properties
in the simulation and the fourier-space properties of the assumed cosmological model, 
I here focus the discussion on a method for running ensembles of simulations that is designed to instead
 maximize the correspondence between simulated real-space clustering
statistics (e.g. $\sigma_8, \xi(r)$) and the real-space properties of the 
assumed cosmological model. Originally proposed by Pen \cite{Pen1997} and 
implemented by Sirko \cite{Sirko2005}\footnote{Once publicly available, the code can still be obtained through \mbox{http://web.archive.org}}, as this method allows the so-called DC mode of
each simulation (in an ensemble of simulations) to vary self-consistently according to the 
clustering power on the scale of the box in much the same way that 
the density within randomly placed boxes in the real universe will 
fluctuate around the mean density. In the early days of fully cosmological
$N$-body simulations  \cite[e.g.][]{Frenk_etal1988} this effect was
sometimes included, albeit in less-sophisticated ways than in \cite{Pen1997} and \cite{Sirko2005}.\footnote{This issue has also been discussed in the context of artificially changing the DC mode of an existing simulation as a way of scaling a simulation completed with a certain set of cosmological parameters to a slightly different model \citep{Tormen_Bertschinger1996,Cole1997,Angulo_White2010,Schneider_etal2011}.}

In the Sirko \cite{Sirko2005} framework the initial power spectrum used with
the Zeldovich \cite{Zeldovich1970} (and by extension 2LPT 
\cite{Bouchet_etal1995,Scoccimarro1998}) approximation
is convolved such that the matter correlation function matches exactly
the linear theory correlation function for $r < L_{\rm{box}} / 2$, while for
$r > L_{\rm{box}} / 2$ the correlation function is set to zero. With this in mind
Sirko refers to this approach as ``$\xi$-sampled'' initial conditions (ICs),
while the standard method is referred to as 
``$P$-sampled'', since by using an unconvolved linear theory power spectrum
with the Zeldovich approximation the initial conditions are instead matched to 
the fourier space clustering statistics. The $\xi$-sampled strategy, 
by matching the correlation function out to $r = L_{\rm{box}}/2$, should avoid 
biases on all real space statistics, since the rms overdensity in spheres, 
$\sigma(R)$, is simply related to the correlation function, and the halo mass 
function to good approximation is only a function of $\sigma(R)$ \cite{Reed_etal2003,Jenkins_etal2001,Tinker_etal2008}. 
Without this convolution these real space statistics become biased (e.g. from 
$P(k) = 0$ for $k \lesssim 2 \pi / L_{\rm{box}}$), as  
discussed by \cite{Pen1997} and \cite{BaglaPrasad2006}. Sirko \cite{Sirko2005} 
presents a set of $\Lambda$CDM simulations with 100$ h^{-1}$ Mpc box sizes that 
indicate that the conventional, $P$-sampled method can give strongly biased results
for the matter correlation function on scales near 1/4th the size of the box,
independently of epoch, while the results of $\xi$-sampled simulations with the
same parameters give much more reasonable matter correlation functions 
on these scales. This conclusion is revisited in \S~\ref{sec:cosmostats},
which argues that if a measurement-bias correction is applied to the $P$-sampled results,
the two methods are consistent.

Although a number of groups have published results using the initial 
conditions code developed by Sirko, which was the among the first include
the 2nd order Lagrangian corrections \citep{Bouchet_etal1995,Scoccimarro1998} 
to the Zeldovich \cite{Zeldovich1970} displacements, the code is 
very seldom used to generate $\xi$-sampled ICs. 
To my knowledge, only Reid et al. \cite{Reid_etal2009} have utilized the code in
this mode, citing the success of convergence tests in \cite{Reid2008}. 
In that study they create mock catalogues from a suite of 42 simulations with 
$L_{\rm{box}} = 558 h^{-1}$ Mpc, and $N = 512^3$ for comparison with SDSS LRG
data \citep{Tegmark_etal2006}. They chose the $\xi$-sampled method for this
task, citing the attractive feature of allowing the DC mode of the 
box to vary, thereby modeling the power spectrum covariance of real surveys 
more realistically. \cite{Reid2008} and Appendix A of \cite{Reid_etal2009}
present a wide variety of convergence tests that explore the effects of 
increasing the resolution with either fixed initial conditions (i.e. with a 
particular randomly sampled value for the DC mode) or for a set of a few 
initial conditions realizations. More recently, \cite{Gnedin_etal2011}
argued that the DC mode should be re-introduced and compared the results
of five $L_{\rm{box}} = 20 h^{-1}$ Mpc, $\Lambda$CDM simulations using the
$\xi$-sampled method to a high-resolution, standard-method simulation with 
$L_{\rm{box}} = 80 h^{-1}$ Mpc, finding good correspondence between the 
results for the variance of the halo mass function. 

This study systematically explores the predictions of the
two different methods using relatively large ensembles of simulations 
(20 unless otherwise noted) and a diverse set of initial conditions. Where the 
results disagree it may be ambiguous which approach is more accurate, 
therefore I focus 
on pure powerlaw models which should evolve self-similarly. This allows
highly-accurate self-consistency checks of the simulation results, since each
output should, in a statistical sense, resemble scaled versions of earlier
and later outputs. These kinds of ``self-similar'' tests were decisive
in confirming the accuracy of the first generation of fully cosmological 
N-body codes \citep{Efstathiou_etal1988}. I also show a few tests where, 
instead of a pure powerlaw, I simulate BAO-inspired initial conditions consistent with
a  configuration space powerlaw times a gaussian bump. Investigated
 in great depth in \cite{Orban_Weinberg2011} using the conventional method, this test 
is self-similar in a different 
sense -- namely that the evolution of the dark matter clustering should only depend 
on the ratio of the scale of
non-linearity to the scale of the BAO. I include these 
initial conditions as another test of the $\xi$-sampled method and as 
a valuable cross-check for the conventional method's predictions for the 
non-linear evolution of the BAO feature. Importantly, these simulations
can explore the shift and broadening of the BAO bump even in
the strongly-clustered regime ($\sigma_8 \gtrsim 1$).

I test these models extensively, focusing on pure powerlaw models with
spectral slopes of $n = -1$, $-1.5$, and $-2$, and on the three models 
explored in \cite{Orban_Weinberg2011} which resemble $n = -0.5$, $-1$, and 
$-1.5$ powerlaws in fourier space. \S~\ref{sec:ICs} gives an overview of
the $\xi$-sampled method. \S~\ref{sec:cosmostats} describes aspects of
 measuring the correlation function in $\xi$-sampled and $P$-sampled 
simulations, including the importance of the integral constraint 
measurement bias which led Sirko \cite{Sirko2005} to believe incorrectly that 
correlation functions in $P$-sampled, $\Lambda$CDM simulations are 
suppressed for $r \gtrsim L_{\rm box}/10$. \S~\ref{sec:xisamppow} describes powerlaw initial conditions
in the $\xi$-sampled context. I compare predictions from the two
methods, showing results for the matter-matter two-point correlation function in \S~\ref{sec:xi}.
In \S~\ref{sec:boxtobox} I investigate results for the variance of the 
correlation function, comparing the results from the two methods to each other
and to expectations from theory.  In \S~\ref{sec:end} I summarize my main conclusions. 
Appendix~\ref{ap:ximatter} presents a simple, independent-of-epoch analytic formula
that, given the box size, can estimate the importance of the integral constraint in 
$\Lambda$CDM simulations.

\section{Overview of the $\xi$-sampled Method}
  \label{sec:ICs}

In the $\xi$-sampled method implemented by \cite{Sirko2005}, the (real space) matter 
correlation function for a given cosmological model is the (usual) fourier transform of
 the power spectrum
\begin{equation}\label{eq:ft}
  \xi(r) = \int \frac{d^3k}{(2 \pi)^3} P(k) \, e^{i \vec{k} \cdot \vec{r}} = \frac{1}{2 \pi^2}\int_0^\infty P(k) \, \frac{\sin k r }{k r} \, k^2 \, {dk}.
\end{equation}
To convolve $P(k)$ such that the simulated $\xi(r)$ is an exact match to 
Eq.~\ref{eq:ft} for $r < L_{\rm{box}} / 2$, but is zero for larger separations, 
one simply fourier transforms $\xi(r)$  while cutting
 off the integral at $L_{\rm{box}} / 2$ since $\xi(r) = 0$ for $r > L_{\rm{box}}/2$,
\begin{equation}\label{eq:preal}
  P_{\rm{real}}(k) = 4 \pi \int_0^{L_{\rm{box}}/2} \xi(r) \, \frac{\sin k r}{k r} \, r^2 \, {dr}.
\end{equation}
I will refer to this result as $P_{\rm{real}}(k)$ to emphasize that this power
spectrum is designed to maintain correspondence with the real space properties
of the cosmological density field.  Importantly, $P_{\rm{real}}(0)$ can be non-zero
even if $P(0) = 0$; this term sets the fluctuations in the DC mode.
In Appendix A of \cite{Sirko2005}, using the subscript ``uni'' to denote 
variables in the model of interest and ``box'' to identify the parameters
of the simulated volume,  these fluctuations are mapped 
self-consistently onto fluctuations in cosmological parameters,
\begin{eqnarray}
H_{0,\rm{box}} = H_{0,\rm{uni}} \frac{1}{1+\phi}\label{eq:hubble}, \\
\Omega_{m,\rm{box}} = \Omega_{m,\rm{uni}} (1+\phi)^2, \\
\Omega_{\Lambda,\rm{box}} = \Omega_{\Lambda,\rm{uni}} (1+\phi)^2, \\
\phi = \frac{5}{6}\frac{\Omega_m}{D(1)}\Delta_0, \label{eq:phi}
\end{eqnarray}
where $\Delta_0$ is a gaussian variable with mean zero and variance 
$P_{\rm{real}}(0)/L_{\rm{box}}^3$ and $D(1)$ is the value of the linear growth 
function at the present epoch. Note that Eq.~\ref{eq:hubble} implies that in
$h^{-1}$ length units the box size of each simulation varies with the value of
$\phi$, whereas in length units without the inverse hubble factor (e.g. Mpc) the box size remains fixed.
Similarly the box integrated mass, $M_{\rm{box}} = \rho_m L_{\rm{box}}^3$, varies 
from box-to-box in $h^{-1} M_\odot$ units, but is fixed in $M_\odot$ units. 

Of crucial importance in deriving 
Eqs.~\ref{eq:hubble}-\ref{eq:phi} is the relationship between the 
scale factor of interest, $a_{\rm{uni}}$, and the corresponding scale factor in 
a particular realization, $a_{\rm{box}}$. In \cite{Sirko2005} this relationship is 
set by an approximate formula which determines $a_{\rm{box}}$ as the epoch where
the age of the universe in the box is the same as the age of the 
unperturbed universe during the epoch of interest,\footnote{\cite{Cole1997} was the first to appreciate that $a_{\rm{box}} \neq a_{\rm{uni}}$ but instead proposed to set $a_{\rm{box}}$ by matching the amplitude of the linear growth function in the perturbed cosmology.}
\begin{equation}
a_{\rm{box}} \approx a_{\rm{uni}} \left( 1 - \frac{1}{3} \frac{D(a_{\rm uni})}{D(1)} \Delta_0 \right). \label{eq:abox}
\end{equation}
\cite{Sirko2005} justified this formula by arguing that the ratio of the average density of the universe to the average density of a given box, 
$\bar{\rho}_{\rm{uni}} / \bar{\rho}_{\rm{box}} = a_{\rm{box}}^3 / a_{\rm{uni}}^3$,
is simply related to the overdensity of the box, which grows
according to the linear theory growth function. Eq.~\ref{eq:abox} can also
be obtained by Taylor expanding the perturbed $H(a_{\rm{box}})$ for small $\phi$
and equating the age of the universe in the box to the age of the 
universe at the epoch of interest.

\subsection{Integration of Particle Trajectories}

Having set up the initial conditions, determined the perturbed cosmological
parameters of a given realization and computed the relevant scale factors, 
$a_{\rm{box}}$, for the epochs of interest, the initial conditions can be 
evolved using any cosmological N-body code. I use the publicly-available
Gadget2 code with no modifications \citep{Springel2005}. As a hybrid 
Tree-based code with a PM grid for large scale forces, Gadget2 is a highly 
scalable N-body code which compares well to other codes used in the literature 
\citep[e.g.][]{Heitmann_etal2010}. Unless otherwise noted I show results
from simulations with $256^3$ particles and a $512^3$ PM grid. Initial
redshifts were set using $\Delta^2(k_{\rm{Ny}}) \lesssim 0.001$ as a 
rule of thumb \citep{Lukic_etal2007}, and the force softening was set to 
1/20th the initial mean interparticle spacing.

\section{Measurements of the Two-Point Correlation Function}
\label{sec:cosmostats}

With ensembles of simulations in the conventional method, the 
measurements of dark matter clustering at a given output,
$a_{\rm{uni}}$, can typically be combined, and the statistical precision
improved, with a simple average.
In $\xi$-sampled simulations this procedure is somewhat more complicated.
For clarity, the Sirko 2005 approach for measuring the matter-matter two-point
correlation function will be described in \S~\ref{sec:sirkoest}, and
then a conceptual subtlety with this formulation will be highlighted
with an alternate derivation in \S~\ref{sec:better}. Following these 
subsections, the integral constraint bias will be discussed in both
the $P$-sampled and $\xi$-sampled contexts.

In what follows $_{,i}$ subscripts are used to distinguish quantities
that change from realization to realization from those without 
$_{,i}$ subscripts that stay fixed. Also, it is helpful to remember that the number of simulation particles in each 
realization is kept fixed and that the box size is fixed in Mpc units, so in 
any box $i$,
\begin{equation}
\bar{n}_{{\rm box},i} L_{{\rm box},i}^3 = \bar{n}_{\rm uni} L_{\rm uni}^3 = N  \label{eq:nL3}
\end{equation}
where $L_{\rm uni}$ is the mean box size of the realizations in comoving Mpc
units ($L_{{\rm box},i}  =  a_{{\rm box},i} L_{\rm uni} / a_{\rm uni}$)
and accordingly both $\bar{n}_{{\rm box},i}$ and $\bar{n}_{\rm uni}$ are in Mpc$^{-3}$ units (instead of $h^3$ Mpc$^{-3}$ units).
Depending on the context, $N$ is either the total number of simulation particles in the 
box or the total number of randomly-selected tracer particles being used to compute the 
correlation function. Both contexts hold $N$ fixed and therefore $\bar{n}_{{\rm box},i}$ 
and $\bar{n}_{\rm uni}$ are simply related,
\begin{equation}
\bar{n}_{{\rm box},i} a_{{\rm box},i}^3 = \bar{n}_{\rm uni} a_{\rm uni}^3. \label{eq:na3}
\end{equation}
This also connects the scale factors to the overdensity,
\begin{equation}
\frac{\bar{n}_{{\rm box},i}}{\bar{n}_{\rm uni}} = \left( \frac{a_{\rm uni}}{a_{{\rm box},i}} \right)^3 \equiv 1 + \frac{D(a_{\rm uni})}{D(1)} \Delta_{0,i}
\end{equation}
which is very similar to the expression in Eq.~\ref{eq:abox}. For brevity, the symbol $\Delta_i \equiv (D(a_{\rm uni}) / D(1)) \Delta_{0,i}$ will frequently be used to denote the 
overdensity of a given box at a particular epoch.

\subsection{Estimation in Sirko 2005}
\label{sec:sirkoest}

In the Sirko 2005 approach, the principal subtlety in calculating the 
mean correlation function from an ensemble of $\xi$-sampled simulations is
simply that the mean number density in each box, $\bar{n}_{{\rm box},i}$, 
deviates from the mean number density, $\bar{n}_{\rm uni}$.

We naturally begin with a correlation function measurement that is totally ignorant of the ``uni'' cosmology. 
Using the overdensity, $\delta = n / \bar{n} - 1$, and the well-known formula for the two-point correlation function, this is
\begin{equation}
\xi_{\, {\rm box},i} (r) = \langle \delta_{{\rm box},i}(\vec{x}) \, \delta_{{\rm box},i}(\vec{x} + \vec{r}) \rangle = \left \langle \left(\frac{n_i(\vec{x})}{\bar{n}_{{\rm box},i}}  -1   \right) \cdot \left( \frac{n_i(\vec{x}+\vec{r})}{\bar{n}_{{\rm box},i}}  -1   \right) \right \rangle \label{eq:xiboxi}
\end{equation}
where the $\langle \, \rangle$ symbols denote an average over the simulation box; $n_i(\vec{x})$ and $n_i(\vec{x}+\vec{r})$ are number densities at different 
positions within the box, $i$. It is straightforward to show that Eq.~\ref{eq:xiboxi} is equivalent to
\begin{equation}
\xi_{\, {\rm box},i} (r) = \frac{\langle n_i (\vec{x}) \, n_i (\vec{x}+\vec{r}) \rangle}{\bar{n}_{{\rm box},i}^2} - 1
\end{equation}
since $\langle n_i (\vec{x}) \rangle = \langle n_i (\vec{x}+\vec{r}) \rangle = \bar{n}_{{\rm box},i}$. The goal now is to find the relation between $\xi_{{\rm box},i}(r)$ and a correlation function measurement in the ``uni'' cosmology,
\begin{equation}
\xi_{{\rm uni},i} (r) = \langle \delta_{\rm uni} (\vec{x}) \, \delta_{\rm uni}(\vec{x} + \vec{r}) \rangle = \left \langle \left(\frac{n_i(\vec{x})}{\bar{n}_{\rm uni}}  -1   \right) \cdot \left( \frac{n_i(\vec{x}+\vec{r})}{\bar{n}_{\rm uni}}  -1   \right) \right \rangle . \label{eq:xiunii}
\end{equation} 
Using Eq.~\ref{eq:na3}, Eq.~\ref{eq:xiunii} can be expanded to become,
\begin{eqnarray}
\xi_{{\rm uni},i}(r) & \displaystyle = \frac{\langle n_i (\vec{x}) \, n_i (\vec{x}+\vec{r}) \rangle}{\bar{n}_{\rm uni}^2} - \frac{\langle n_i (\vec{x}) \rangle }{\bar{n}_{\rm uni}} - \frac{\langle n_i (\vec{x}+\vec{r}) \rangle }{\bar{n}_{\rm uni}}+ 1 \nonumber  \\
& \displaystyle = \left( \frac{a_{\rm uni}}{a_{{\rm box},i}} \right)^6 \frac{\langle n_i (\vec{x} ) \, n_i (\vec{x}+\vec{r}) \rangle}{\bar{n}_{{\rm box},i}^2} - 2 \left(\frac{a_{\rm uni}}{a_{{\rm box},i}}\right)^3  + 1. \label{eq:xiunii2}
\end{eqnarray}
Combining Eqs.~\ref{eq:xiboxi} \& \ref{eq:xiunii2} we obtain,
\begin{equation}
\xi_{{\rm uni},i}(r) = \left( \frac{a_{\rm uni}}{a_{{\rm box},i}} \right)^6 \left(\xi_{{\rm box},i}(r) + 1   \right) - 2 \left(\frac{a_{\rm uni}}{a_{{\rm box},i}}\right)^3  + 1. \label{eq:xiunii3}
\end{equation}
which is equivalent to Eq.~25 from Sirko 2005. In the final averaging, $\xi_{{\rm uni},i}(r)$ in Eq.~\ref{eq:xiunii3} is weighted by $w_i = (a_{{\rm box},i} / a_{\rm uni})^3$ to ensure that
boxes with larger volumes receive higher weight. Unless otherwise noted Eq.~\ref{eq:xiunii3} is used with the weighting just mentioned in
calculations of the two-point correlation function in $\xi$-sampled simulation ensembles.

\subsection{Subtleties of Eq.~\ref{eq:xiunii3}: Survey-like versus ``better informed'' estimators}
\label{sec:better}

To highlight the subtleties of Eq.~\ref{eq:xiunii3}, let us re-derive the expression in a different way. 
The two-point correlation function can be equivalently defined as the joint probability, $\delta P$,
to find a particle in volume, $dV_1$, and another particle, at some distance, $r$, in the volume $dV_2$,
\begin{equation}
\delta P = \bar{n}^2 dV_1 dV_2 (1 + \xi(r)). \label{eq:2pt}
\end{equation}
For a given realization, one of these volume elements integrate to the volume of
the simulation box, $L_{{\rm box},i}^3$, while the other volume is integrated
over a radial shell, $V_{\rm shell}$. 
For the correlation function of
an individual box, $\xi_{{\rm box},i}(r)$, for which $\bar{n} = \bar{n}_{{\rm box},i}$,
this yields an expression for the total number of pairs in the box
within a given radial separation,
\begin{equation}
N_{p,i} (r, \Delta r) = \frac{1}{2}\bar{n}_{{\rm box},i}^2 L_{{\rm box},i}^3 V_{\rm shell} (1 + \xi_{{\rm box},i}(r)) \label{eq:pairbox}
\end{equation}
where the $1/2$ factor avoids the double counting of pairs. 
The above expression is useful as an algorithm for measuring $\xi_{{\rm box},i}(r)$ from counting the number of 
pairs at various separations in a given simulation box.

There are two ways of converting $\xi_{{\rm box},i} (r)$ in Eq.~\ref{eq:pairbox} into 
a correlation function measurement in the ``uni'' cosmology. Most simply, one can 
define $\xi_{{\rm uni},i}(r)$ according to Eq.~\ref{eq:2pt} using the ``uni'' number 
density for $\bar{n}$ and appreciating that the correlation function measurement
is over the volume of the box for a specific realization, $L_{{\rm box},i}^3$,
\begin{equation}
N_{p,i} (r, \Delta r) = \frac{1}{2}\bar{n}_{\rm uni}^2 L_{{\rm box},i}^3 V_{\rm shell} (1 + \xi_{{\rm uni},i}(r)). \label{eq:pairuni}
\end{equation}
This leads to the conclusion that
\begin{equation}
  \xi_{{\rm uni},i}(r) =  \frac{\bar{n}_{{\rm box},i}^2}{\bar{n}_{\rm uni}^2}  \left( \xi_{{\rm box},i}(r) + 1 \right) - 1 = \left( \frac{a_{\rm uni}}{a_{{\rm box},i}} \right)^6 \left( \xi_{{\rm box},i}(r) + 1 \right) - 1. \label{eq:xiunii4}
\end{equation}
The remarkable consequence of assuming Eq.~\ref{eq:xiunii4} is that even if the particle distribution in the simulation volume is completely
uncorrelated $(\xi_{{\rm box},i}(r) \rightarrow 0)$, the correlation function in the ``uni'' cosmology can still be non-zero since, in that case,
\begin{equation} 
\xi_{{\rm uni},i}(r) = \left(\frac{a_{\rm uni}} { a_{{\rm box},i}} \right)^6 - 1 \approx 2 \Delta_i.  \label{eq:uncorr}
\end{equation}
Importantly this result remains after volumetric weighting is applied to $\xi_{{\rm uni},i}(r)$.

From the non-zero result of Eq.~\ref{eq:uncorr} it is clear that Eq.~\ref{eq:xiunii4} is a 
survey-like approach to measuring the correlation function in the simulation ensemble 
in the sense that the measurement
 knows about the volume of the box but it does not know the true overdensity
of the box. This ignorance is transferred to $\xi_{{\rm uni},i}(r)$ and it is only in averaging over 
many simulations that the mean of the $\Delta_i$ values will be close to zero and a precise measurement of
the mean correlation function can be made. This is very much like surveys where, in principle,
one would benefit from perfectly knowing the overdensity of a particular subvolume which would be 
useful for measuring the correlation function. Perfect knowledge of the overdensity would help 
determine how much of a measured excess (or decrement) of pairs 
in a subvolume reflects the the true non-linear correlation function 
and how much of the excess (or decrement) reflects a difference between
the mean density of the subvolume and the mean density of the universe.
However, in practice, the overdensity of a particular subvolume in a survey is uncertain at some level and this uncertainty must 
be taken into account in estimating the errors on the clustering measurement.

A more-sophisticated (a.k.a. ``better-informed'') approach to connecting $\xi_{{\rm box},i}(r)$ and $\xi_{{\rm uni},i}(r)$ 
is therefore to use the overdensity information, as just described, to compare the 
measured number of pairs, $N_{p,i}(r,\Delta r)$, to a ``better-informed'' expectation of the number 
of random pairs for that simulation volume. To do this one can introduce a correlation function 
offset, denoted by $\xi_{\, \delta,i}$, that will make this adjustment, 
\begin{equation}
N_{p,i} (r, \Delta r) = \frac{1}{2}\bar{n}_{\rm uni}^2 L_{{\rm box},i}^3 V_{\rm shell} (1 + \xi_{\, \delta,i} + \xi_{{\rm uni},i}(r)). \label{eq:pairuni2}
\end{equation}
At large separations, or in a hypothetical situation where the clustering in each box 
is totally uncorrelated, then $\xi_{{\rm box},i}(r) \rightarrow 0$ and we can define 
$\xi_{{\rm uni},i}(r)$ so that by fiat in each box $\xi_{{\rm uni},i}(r) \rightarrow 0$ and the box-to-box fluctuations in overdensity
are entirely captured by $\xi_{\delta,i}$. This implies
\begin{equation}
1 + \xi_{\, \delta,i} = \frac{\frac{1}{2} \bar{n}_{{\rm box},i}^2 L_{{\rm box},i}^3 V_{\rm shell} }{\frac{1}{2} \bar{n}_{\rm uni}^2 L_{{\rm box},i}^3 V_{\rm shell}} =\left( \frac{a_{\rm uni}}{a_{{\rm box},i}} \right)^6  \approx 1 + 2 \Delta_i .
\end{equation}
or just
\begin{equation}
\xi_{\delta,i} \approx 2 \Delta_i.
\end{equation}
Solving for $\xi_{{\rm uni},i}(r)$ in Eq.~\ref{eq:pairuni2}, the ``better informed'' estimator becomes
\begin{eqnarray}
\xi_{{\rm uni},i}(r) & = &\displaystyle   \frac{N_{p,i}(r,\Delta r)}{\frac{1}{2} \bar{n}_{\rm uni}^2 L_{{\rm box},i}^3 V_{\rm shell}} - 1 - 2 \Delta_i \nonumber \\
& = &\displaystyle  \left( \frac{a_{\rm uni}}{a_{{\rm box},i}} \right)^6 \frac{N_{p,i}(r,\Delta r)}{\frac{1}{2} \bar{n}_{{\rm box},i}^2 L_{{\rm box},i}^3 V_{\rm shell}} - 1 - 2 \left( \left( \frac{a_{\rm uni}}{a_{{\rm box},i}} \right)^3 -1 \right) \nonumber \\
& = & \displaystyle  \left( \frac{a_{\rm uni}}{a_{{\rm box},i}} \right)^6 (\xi_{{\rm box},i}(r) + 1)  - 2  \left( \frac{a_{\rm uni}}{a_{{\rm box},i}} \right)^3 + 1 
\end{eqnarray}
which is identical to the result Sirko derived (Eq.~\ref{eq:xiunii3} in this work). 
Sirko's estimator therefore implicitly uses the knowledge of the overdensity in 
each box to improve the correlation function estimate. Parenthetically, note that as in Eq.~\ref{eq:xiunii3} and in Sirko \cite{Sirko2005}
the above expression for $\xi_{{\rm uni},i}(r)$ must be volumetrically weighted by $w_i = (a_{{\rm box},i} / a_{\rm uni} )^3$ when 
averaging over all realizations.

Interestingly, this ``better-informed'' estimator is not unlike correlation function measurements in 
conventional, $P$-sampled simulations.  
Since the density of finite volumes in the real universe fluctuates around the mean, 
arguably one should account for this source of uncertainty in the error bars of 
a given correlation function measurement from a $P$-sampled simulation. But instead, rather than degrade
the error on the mean correlation function, one naturally uses the extra information that 
the overdensity of a given $P$-sampled simulation is always zero, regardless 
of the box size, to inform the expectation for the number of random pairs. 
Thus for $P$-sampled simulations $\bar{n}_{{\rm box},i}$ is always equal to $\bar{n}_{\rm uni}$ (in general and in Eq.~\ref{eq:pairuni2})
and consequently {\it it is perfectly known} that $\Delta_i = 0$ (i.e. $\xi_{\delta,i} = 0$) for all realizations.
In this sense, correlation function measurements in $P$-sampled simulations are
also performed with a ``better informed'' estimator without any extra effort.

\subsection{Integral-Constraint Bias in $P$-sampled Simulations}
\label{sec:intbiassec}

An important but sometimes neglected
measurement bias that affects correlation function estimation is an integral
constraint that arises from the fact that summing over the number of pairs
in the volume must naturally yield $\frac{1}{2} N^2$ where $N$ is the number
of randomly selected tracer particles. This issue has been identified
by other authors (e.g. \cite{Bernardeau_etal2002}) and it is
is entirely orthogonal to the question of which estimator 
\citep[][etc.]{Davis_Peebles1977,Landy_Szalay1993} converges most rapidly to the true $\xi(r)$ in the
presence of Poisson noise. Orban \& Weinberg \cite[][Appendix B]{Orban_Weinberg2011} outline an
approach for correcting the correlation function measurement.
Appendix~\ref{ap:ximatter} demonstrates that for $\Lambda$CDM simulations with large boxes
($L_{\rm box} \gtrsim 2 \, h^{-1}$Gpc) the integral constraint is a minor issue. For significantly smaller
boxes this is an important concern.  Notably, \cite{Sirko2005} presented
simulations with $L_{\rm{box}} = 50-100 \, h^{-1}$Mpc without any kind of correction
for this effect. The present section will discuss the integral constraint 
bias in $P$-sampled simulations. This subtlety is also relevant to $\xi$-sampled 
simulations. The next subsection will discuss how the $\xi$-sampled approach
using Eq.~\ref{eq:xiunii3} as in Sirko \cite{Sirko2005}, includes a correction for the problem.

Since the notation in this section differs slightly from that in Orban \& Weinberg \cite[][Appendix B]{Orban_Weinberg2011},
a brief re-derivation of that result will help explain the problem.
For $P$-sampled simulations, the number of pairs in a given radial bin is given by\footnote{Since the box size is fixed in $P$-sampled simulations and the overdensity in each realization is zero, the notation in this subsection uses $L_{\rm box}$ to denote the usual, unchanging box size in comoving coordinates and $\bar{n}$ as the conventionally-defined mean number density in the box which is likewise unchanging.}
\begin{equation}
N_{p,i} (r, \Delta r) = \frac{1}{2}\bar{n}^2 L_{\rm box}^3 V_{\rm shell} (1 + \xi_{{\rm uni},i}(r)). \label{eq:pairuni3}
\end{equation}
If integrated over the entire box this expression becomes
\begin{equation}
\int N_{p,i} (r , \Delta r) = \frac{1}{2}\bar{n}^2 L_{\rm box}^3 \int_0^{R_{\rm box}} ( 1 + \xi_{{\rm uni},i} (r)) \, 4 \pi r^2 dr = \frac{N^2}{2} \label{eq:pairconstraint}
\end{equation}
where $\frac{4}{3} \pi R_{\rm box}^3 \equiv L_{\rm box}^3$, implying that $R_{\rm box} = (4 \pi / 3)^{-1/3} L_{\rm box} \approx L_{\rm box} / 1.61$. Note that $\bar{n} L_{\rm box}^3~=~N$, so Eq.~\ref{eq:pairconstraint} becomes
\begin{equation}
\bar{n} \left[ \frac{4}{3} \pi R_{\rm box}^3 + 4 \pi \int_0^{R_{\rm box}} \xi_{{\rm uni},i}(r) r^2 dr \right] = N
\end{equation}
and since $\bar{n} \, \frac{4}{3} \pi R_{\rm box}^3 = \bar{n} \, L_{\rm box}^3 = N$, a {\it measurement constraint} is imposed on $\xi_{{\rm uni},i}(r)$,
\begin{equation}
\int_0^{R_{\rm box} = L_{\rm box} / 1.61} \xi_{{\rm uni},i}(r) \, r^2 dr = 0. \label{eq:constraint}
\end{equation}
\begin{figure}
\centerline{\epsfig{file=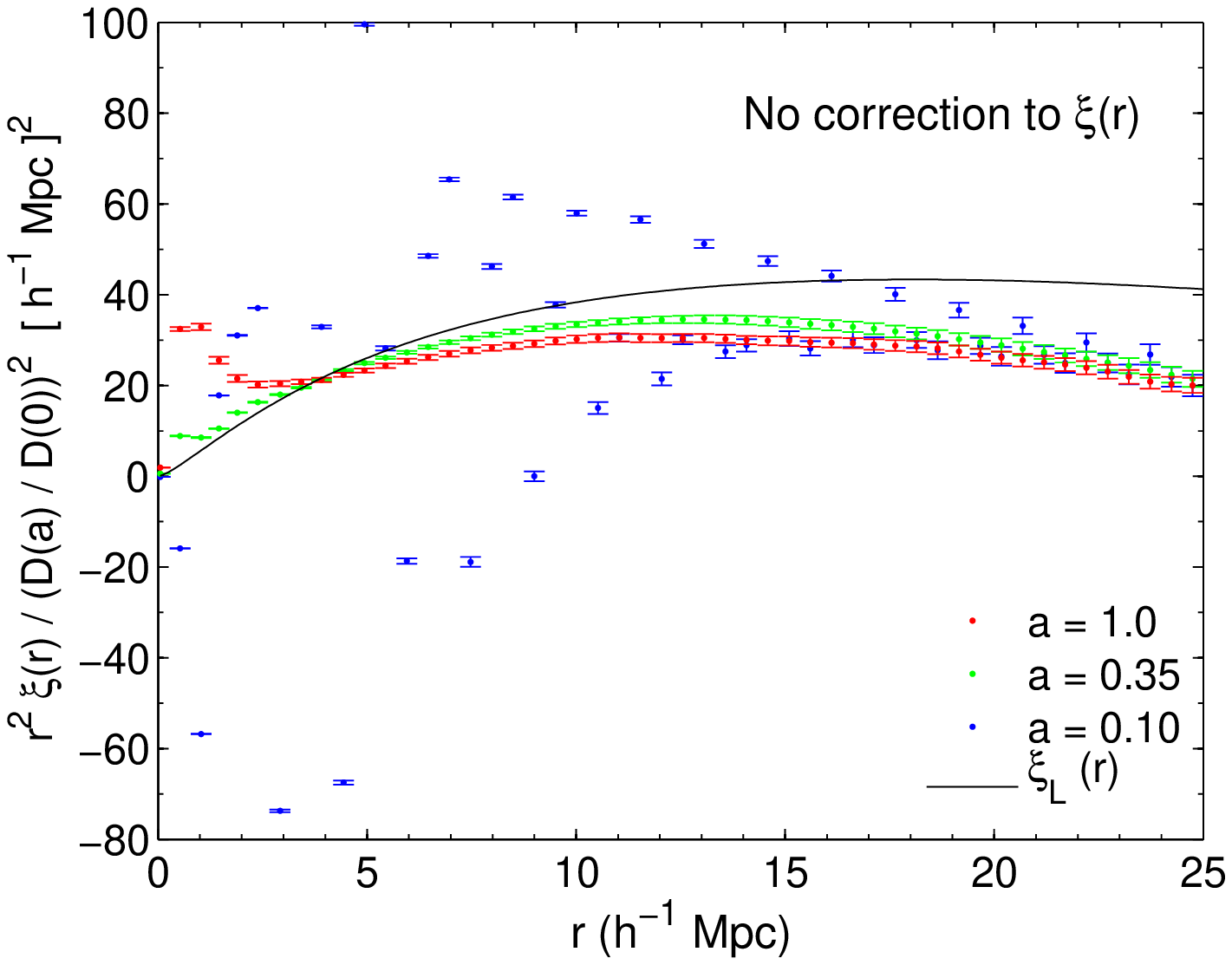, width = 3.15in}\epsfig{file=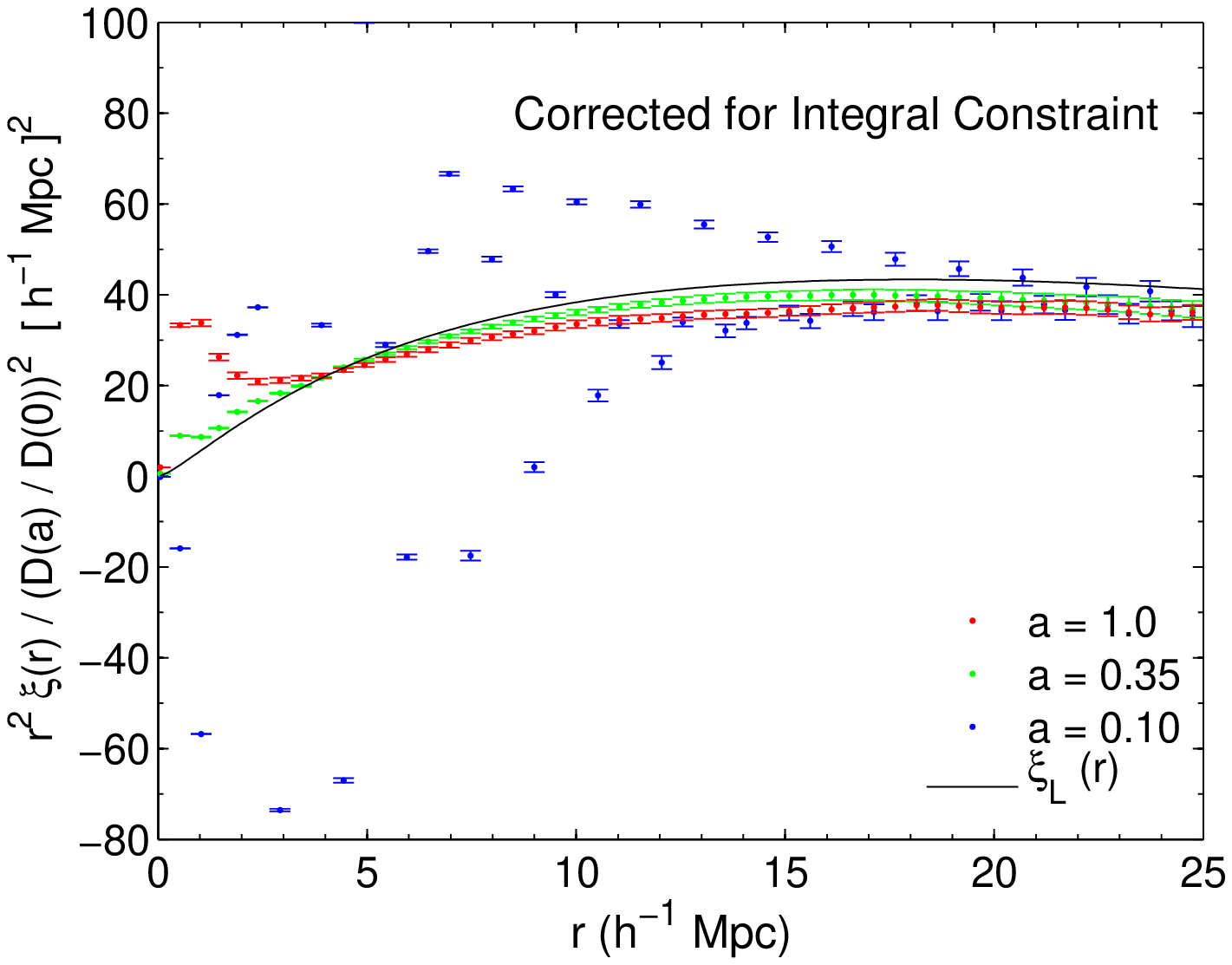, width = 3.15in}}
\vspace{-0.1in}
\caption{ Matter correlation function results from a $\Lambda$CDM 
ensemble of simulations ($L_{\rm{box}} = 100 \, h^{-1}$Mpc, $N = 64^3$, 100
realizations) using standard ($P$-sampled) ICs. The left panel shows the $\xi(r)$
measurements from this simulation set without applying the integral constraint
correction. The right panel shows the results from
including the correction derived in \cite[][Appendix B]{Orban_Weinberg2011}. Error bars show the error
on the mean. Although the earliest output ($a = 0.1$, shown in blue) 
is severely affected by transients from the initial conditions, it is included 
for comparison to Sirko \cite{Sirko2005}, Fig.~9.
}\label{fig:xicorr}
\end{figure}
Our reproduction of the $P$-sampled $\Lambda$CDM simulations presented in Fig.~9 of Sirko 2005 \cite{Sirko2005}, 
shown here in the left panel of Fig.~\ref{fig:xicorr}, indicates that this measurement bias is quite important for the 
$L_{\rm box} = 100 \, h^{-1}$Mpc simulations they present, suppressing the correlation function
at 1/4th the scale of the box by almost a factor of two and causing a severe disagreement
with the linear theory correlation function for $r \sim 20-25 \, h^{-1}$Mpc despite the 
fact that $\xi_L (r) \ll 1$ on these scales.

To correct for this measurement bias, following the approach used in Orban \& Weinberg \cite{Orban_Weinberg2011},
one defines 
\begin{equation}
\xi_{{\rm uni},i} (r) = \xi_{{\rm uni,true},i}(r) + \xi_{\rm bias}
\end{equation}
where $\xi_{\rm bias}$ is a radially-independent term and $\xi_{{\rm uni,true},i}(r)$ is 
understood to be the correlation function of the box without the integral-constraint bias.
Using Eq.~\ref{eq:constraint}, $\xi_{\rm bias}$ can be solved for, giving 
\begin{equation}
\xi_{\rm bias} =  -\frac{3 }{ R_{\rm box}^3} \int_0^{R_{\rm box} = L_{\rm box} / 1.61} \xi_{{\rm uni,true},i}(r) \, r^2 dr \approx -\frac{3 }{ R_{\rm box}^3} \int_0^{R_{\rm box} = L_{\rm box}/1.61} \xi_L (r) \, r^2 dr   \label{eq:xibias}
\end{equation} 
where the integral over $\xi_{{\rm uni,true},i}(r)$, which is weighted heavily towards large scales, has been well approximated using linear theory.
The corrected estimator for the correlation function is therefore
\begin{equation}
\xi_{{\rm uni,true},i}(r) = \xi_{{\rm uni},i}(r) - \xi_{\rm bias} = \xi_{{\rm uni},i}(r) + \frac{3 }{ R_{\rm box}^3} \int_0^{R_{\rm box} = L_{\rm box}/1.61} \xi_L (r) \, r^2 dr  .\label{eq:xitrue}
\end{equation}
This result is identical to the prescription presented in Orban \& Weinberg \cite{Orban_Weinberg2011}. 
Results from using the integral-constraint corrected estimator are presented in the right panel of Fig.~\ref{fig:xicorr}.
For separations of $r \sim 20 -25 h^{-1}$ Mpc the amplitude of the correlation function is nearly a 
factor of two higher at all epochs, which agrees much better with the linear theory correlation function
on these scales as would be expected. {\it Therefore the conclusion in Sirko 2005 that $P$-sampled simulations suppress
the correlation function for separations approaching the box scale is found to 
stem from an overlooked integral-constraint correction. } Importantly, as is clear from Fig.~\ref{fig:xicorr},
the integral constraint correction matters for separations as small as $r \sim 10 \, h^{-1}$ Mpc $ \sim L_{\rm box} / 10$
or perhaps slightly {\it smaller}. While most practitioners would regard clustering measurements at 
separations of $r \sim L_{\rm box} / 4$ or $r \sim L_{\rm box} / 5$ in a simulation volume to be 
too large compared to the scale of the box to be trustworthy, it should be received with some amount of surprise
that clustering measurements at separations as small as $r \sim L_{\rm box} / 10$ are significantly biased, independently-of-epoch
in conventional, $P$-sampled simulations. Thankfully, correlation function measurements at these scales can be corrected 
using Eq.~\ref{eq:xitrue} without re-running the simulation and Appendix~\ref{ap:ximatter} provides a useful independent-of-epoch 
formula for $\Lambda$CDM simulations that can estimate this bias at the BAO scale given the size of the simulation box.

\subsection{Integral-Constraint Bias in $\xi$-sampled Simulations}
\label{sec:intbiasxi}

Returning to the conclusions of Sirko \cite{Sirko2005} one may ask why the $\xi$-sampled results 
in \cite{Sirko2005} agreed so well with linear
theory approaching the box scale if Sirko did not also apply an integral constraint correction to the 
$\xi$-sampled correlation function measurements? The answer is that Eq.~\ref{eq:xiunii3} (which is what Sirko
used) includes, in its average, a term very much like the integral constraint
correction. At large enough separations in the simulation box, the particle distribution
will be approximately uncorrelated ($\xi_{{\rm box},i}(r) \approx 0)$. 
On these scales Eq.~\ref{eq:xiunii3} gives
\begin{equation}
\xi_{{\rm uni},i}(r) \approx \left( \frac{a_{\rm uni}} { a_{{\rm box},i} }\right)^6 - 2 \left( \frac{a_{\rm uni}}{a_{{\rm box},i}} \right)^3 + 1 = \left( 1 + \Delta_i \right)^2 - 2 \left( 1 + \Delta_i \right) + 1 = \Delta_i^2 \label{eq:Delta2}
\end{equation}
with $\Delta_i \equiv \frac{D(a_{\rm uni})}{D(1)} \Delta_{0,i}$ as used elsewhere. Taking the average over many realizations $i$,
\begin{equation}
\langle \xi_{{\rm uni},i}(r) \rangle = \langle \Delta_i^2  \rangle = \frac{P_{\rm real}(0)}{L_{\rm box}^3} \frac{D^2(a_{\rm uni})}{D^2(1)}= \frac{4 \pi}{L_{\rm box}^3} \int_0^{L_{\rm box}/2} \xi_L (r) \, r^2 {dr}
\end{equation}
While not exact, the term $P_{\rm real}(0) / L_{\rm box}^3$ is {\it very similar} to the $\xi_{\rm bias}$ correction term 
in Eq.~\ref{eq:xibias} \& \ref{eq:xitrue}. It is this term that does the work, so to speak,
of correcting for the integral constraint in the $\xi$-sampled simulations in \cite{Sirko2005}. 
This is why Sirko concluded that the $\xi$-sampled method matched well with linear theory
on scales approaching the box without an explicit correction term. If this $\Delta_i^2$ term 
had not emerged (as it does in Eq.~\ref{eq:Delta2}, essentially by accident) the $\xi$-sampled correlation 
function results would have been suppressed much like the $P$-sampled result shown in the left panel of Fig.~\ref{fig:xicorr}. But since the term does appear there is no need to explicitly correct 
for the integral constraint bias in $\xi$-sampled simulations. Instead the
correction is understood to be built into Eq.~\ref{eq:xiunii3}, which is 
the formula employed in all the $\xi$-sampled correlation function measurements
presented here.

\section{$\xi$-sampled ICs with Powerlaw Models}
\label{sec:xisamppow}

\begin{figure}
\centerline{\epsfig{file=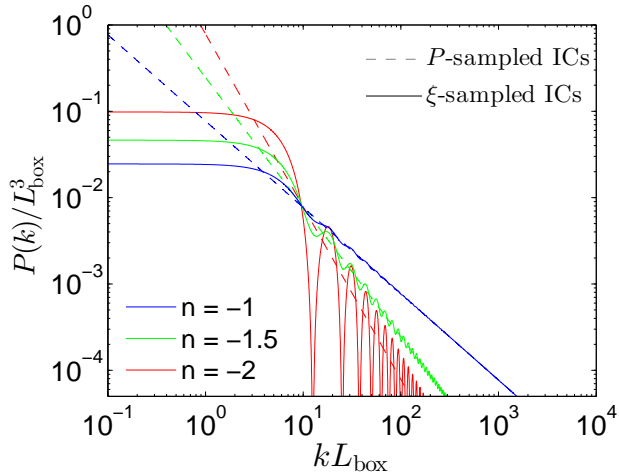, width = 3.5in}}
\vspace{-0.12in}
\caption{ A comparison of $P$-sampled and $\xi$-sampled pure powerlaw
models. $\xi$-sampled power spectra are computed from Eq.~\ref{eq:preal}
and used to generate initial conditions for that method. $r_0 / L_{\rm{box}} = 1/16$ is chosen to set the overall amplitude of each model. To compare with $\xi$-sampled spectra for $\Lambda$CDM initial conditions see Fig.~2 of \cite{Sirko2005}.
}\label{fig:pkcompare}
\end{figure}

For powerlaw models, where $P(k) = A a^2 k^n$, the task of computing 
Eq.~\ref{eq:preal} is made substantially easier  
because an exact analytic solution for $\xi_L (r)$ is known in this case,
\begin{equation}
\xi_L (r) = \left( \frac{r_0}{r} \right)^{n+3}, \, A \, a^2 = \frac{2 \pi^2 \, r_0^{n+3} \, \, (2 + n)}{\Gamma(3+n) \sin((2+n) \pi / 2)} ~, \label{eq:powfac} 
\end{equation}
\citep{Peebles1980}. Eq.~\ref{eq:preal} therefore becomes\footnote{At the early epochs where these 
initial conditions are determined $L_{{\rm box},i} \approx L_{\rm uni}$ to very good approximation.
Therefore to unburden the notation in this section and in a few other places where this approximation is valid
 I opt to use $L_{\rm box}$ (instead of $L_{\rm uni}$) as a more universally recognizable symbol 
for the size of the simulation box. Later, in presenting results from evolved simulations, $\langle L_{{\rm box},i} \rangle$ 
will often be used to denote the average box size instead of the equivalent, $L_{\rm uni}$, to emphasize that
the box size varies from one realization to the next.}
\begin{equation}
  P_{\rm{real}}(k) = 4 \pi r_0^{n+3} \int_0^{L_{\rm box}/2} r^{-(n+1)} \, \frac{\sin k r }{k r} \,  {dr}. \label{eq:ppow}
\end{equation}
Eq.~\ref{eq:ppow} can be used straightforwardly to express the DC power,
\begin{eqnarray}
\displaystyle P_{\rm{real}}(0) \, = & \displaystyle 4 \pi r_0^{n+3} \int_0^{L_{\rm box }/2} r^{-(n+1)} \, {dr} \\
= & \displaystyle \frac{2^{n+2} \pi}{-n} \left( \frac{r_0}{L_{\rm box}} \right)^{n+3} L_{\rm box}^3.  \label{eq:dcpow}
\end{eqnarray}
Analytic and special-function solutions to Eq.~\ref{eq:ppow} exist for certain 
powerlaws. In
this study I am interested in $n = -1$, $-1.5$ and $-2$ which can be 
expressed by
\begin{eqnarray}
P_{\rm{real}, n = -1}(k) = & 4 \pi r_0^2 \, \rm{Si}(k L_{\rm box}/2)\, k^{-1}, \\
P_{\rm{real}, n = -1.5}(k) = & 2^{5/2} \pi^{3/2} r_0^{3/2} \,  \rm{S}( \sqrt{k L_{\rm box}}/\sqrt{\pi}) k^{-1.5}, \\
P_{\rm{real}, n = -2}(k) = & 8 \pi r_0 \, \sin^2(k L_{\rm box}/4) \, k^{-2},
\end{eqnarray}
where $\rm{Si}(x)$ is the sine integral and S($x$) is a Fresnel integral. These 
formulae can be very useful for generating accurate initial conditions, 
especially for steep power spectra. I show these power spectra in 
Fig.~\ref{fig:pkcompare}, fixing $r_0 / L_{\rm box} = 1/16$ to set the relative 
amplitudes. Notice that steeper powerlaws have larger DC power, easily
seen on the plot as the asymptotic value of $P_{\rm{real}}(k) / L_{\rm{box}}^3$ 
as $k \rightarrow 0$. Noticing that $P(k)$ does not go to zero at small $k$ 
for the $P$-sampled powerlaws, one might be concerned that these models are 
unphysical. However, despite the high levels of large scale clustering power
the rms overdensity in spheres and other statistics can remain finite for 
$n > -3$. It so happens that a close inspection of Eq.~\ref{eq:ppow} reveals
that $P_{\rm real} (k)$ is finite and positive (or equal to zero) for all $k$ only 
if $n \geq -2$. It is unclear how to circumvent this issue to simulate 
steeper power 
spectra. For $\Lambda$CDM initial conditions this limitation translates into 
assuming $L_{\rm box}  \gtrsim \, 2.5 h^{-1}$ Mpc to avoid $P_{\rm real} (k) < 0$
because the effective slope of the $\Lambda$CDM correlation function at small 
scales, using $\xi_L (r) = (r_0 / r)^{n_{\rm eff}(r) + 3}$, is $n_{\rm eff} \lesssim -2$ for 
$r \lesssim 2.5 \, h^{-1}$ Mpc.

\subsection{Scale free?}

Although pure powerlaw models are often referred to in the literature as 
``scale free,'' since $P(k) = A k^n$ is featureless, the $\xi$-sampled 
initial power spectra shown in Fig.~\ref{fig:pkcompare} clearly depend
on the choice of $L_{\rm box}$. In practice, these oscillatory features die
away in simulations and the effect of the box size is merely to 
change the variance of the DC mode (which is set by $P_{\rm{real}}(0)/L_{\rm box}^3$).

Since dark energy introduces a new scale into the problem (e.g. the age of the universe when $\rho_m = \rho_\Lambda$), I consider only 
$\Omega_{m,\rm{uni}} = 1.0, \Omega_{\Lambda,\rm{uni}} = 0, \Omega_{k,\rm{uni}} =0$ so as to 
keep the simulations as ``scale free'' as possible and allow the 
self-similar tests discussed in the next section. In the Zeldovich
\cite{Zeldovich1970} and adhesion \citep{Gurbatov_etal1989,Weinberg_Gunn1990}
approximations (as in linear theory), the effect of dark energy on
structure formation is entirely captured by changing the linear theory 
growth function. \cite{Nusser_Colberg1998} and \cite{Zheng_etal2002} 
convincingly argue that this approximation is remarkably accurate even in
the non-linear regime -- the second order effect of dark energy is
relatively small. Therefore the results 
of my $\Omega_m = 1$ tests should still be quite relevant to studies that
include a dark energy component.

As one final comment on the scale-free nature of my simulations, throughout
I adopt, as a time variable,
\begin{equation}
\frac{a}{a_*} =  \left( \frac{k_{\rm{box}}}{k_{\rm{NL}}} \right)^{(n+3)/2} ~, \label{eq:timevar}
\end{equation}
where $k_{\rm{box}} \equiv 2 \pi / L_{\rm{box}}$ and $k_{\rm{NL}}$ is defined by the 
dimensionless linear theory power spectrum, $\Delta_{L}^2(k_{\rm{NL}}) \equiv 1$.
The scale-free nature of the simulations demands that only the ratio 
$a / a_*$ is meaningful (e.g. as the square root of the dimensionless power
on the scale of the box) and so $a_*$ is defined implicitly in the definitions already 
given.
Eq.~\ref{eq:timevar} is also simply related to the $\sigma_{\rm{miss}}$ formula of 
\cite{Smith_etal2003}, which quantifies the missing power on the scale of 
the box in $P$-sampled simulations as another choice for a time variable. 
I adopt Eq.~\ref{eq:timevar} for ease
of comparison with \cite{Widrow_etal09} and because the $\sigma_{\rm{miss}}$
formula in \cite{Smith_etal2003} would be inappropriately applied to 
$\xi$-sampled simulations, which have a turnoff in $P_{\rm real} (k)$ near the box
scale (Fig.~\ref{fig:pkcompare}).

\begin{figure}
\centerline{\epsfig{file=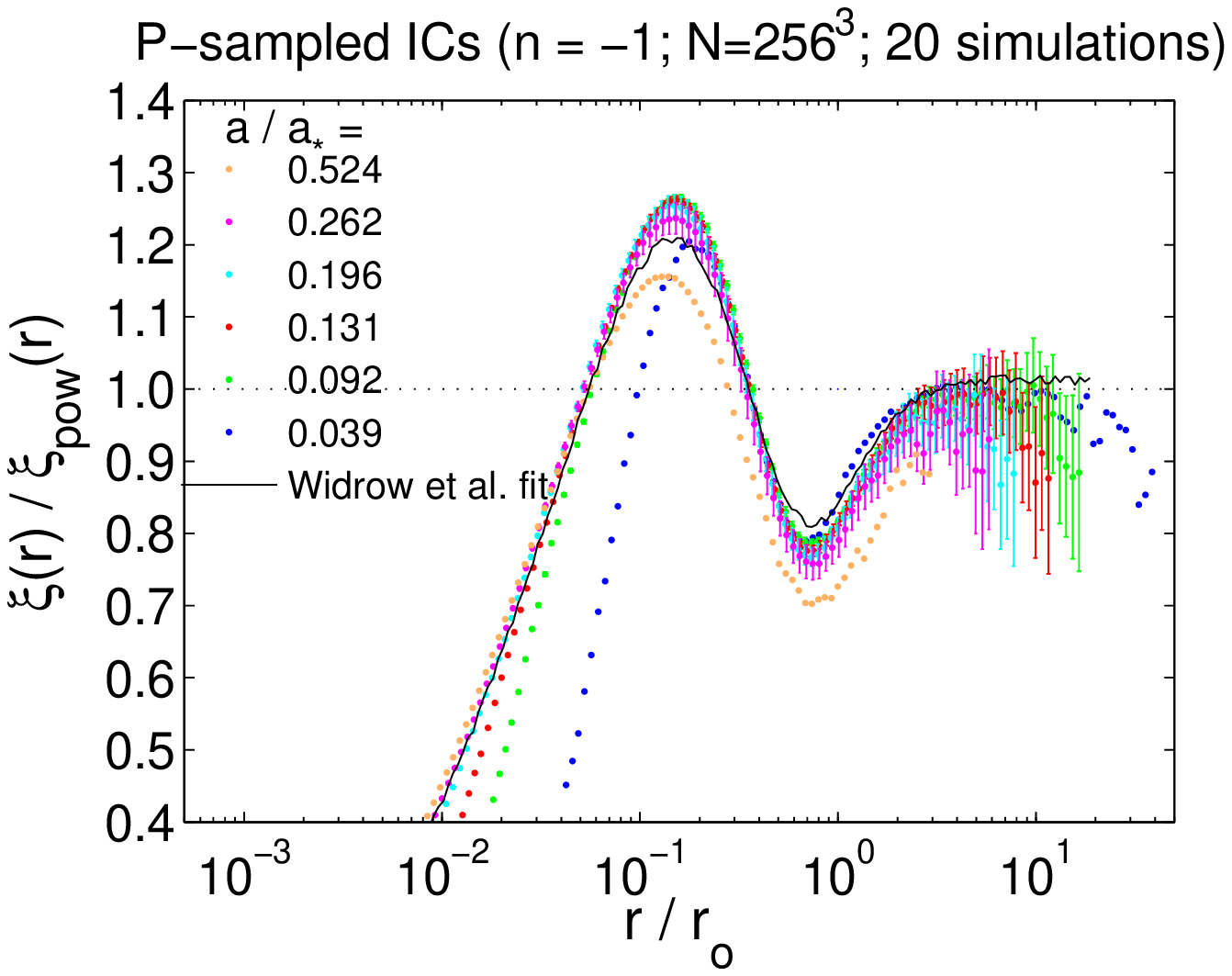, width = 3in}\epsfig{file=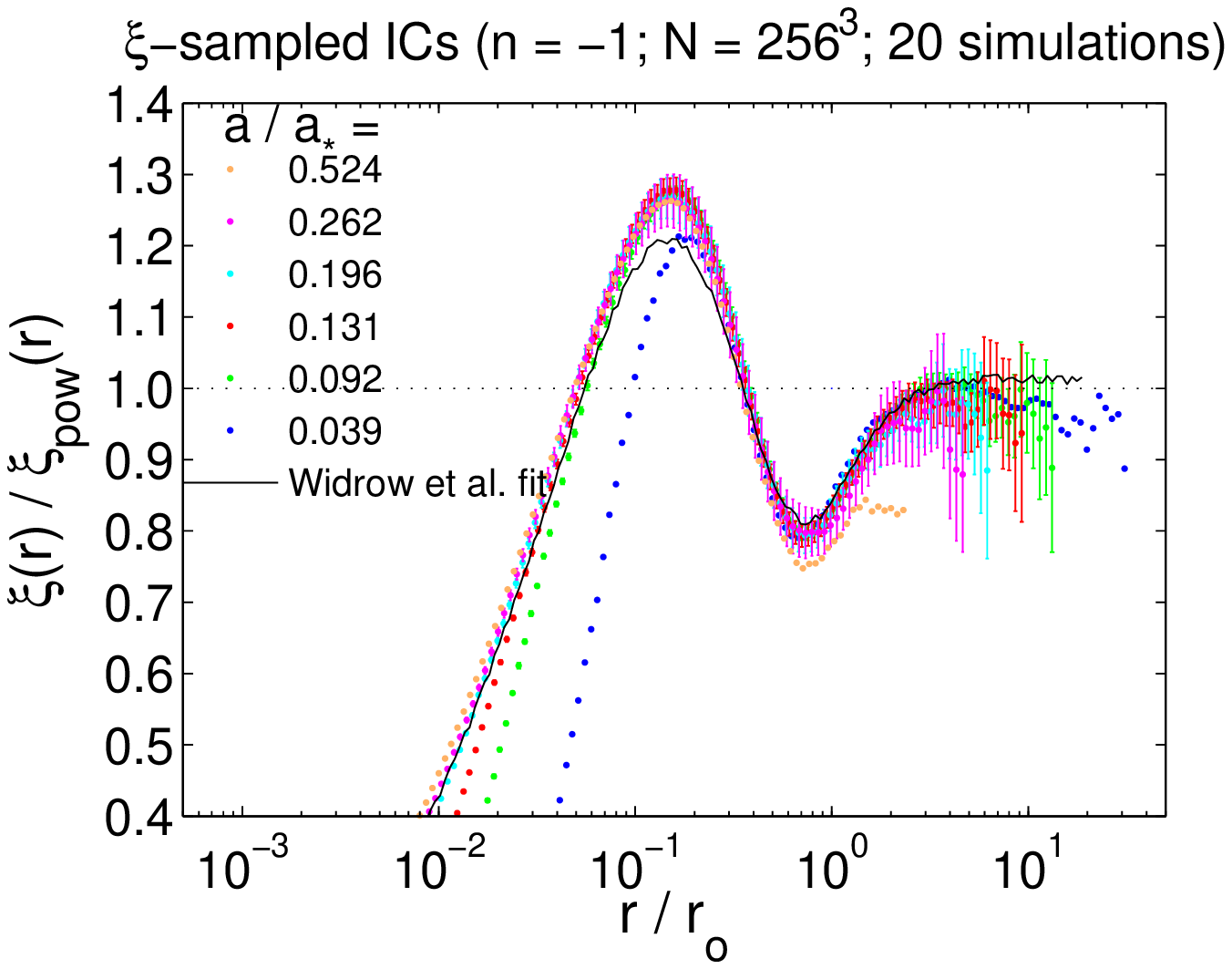, width = 3in}}
\centerline{\epsfig{file=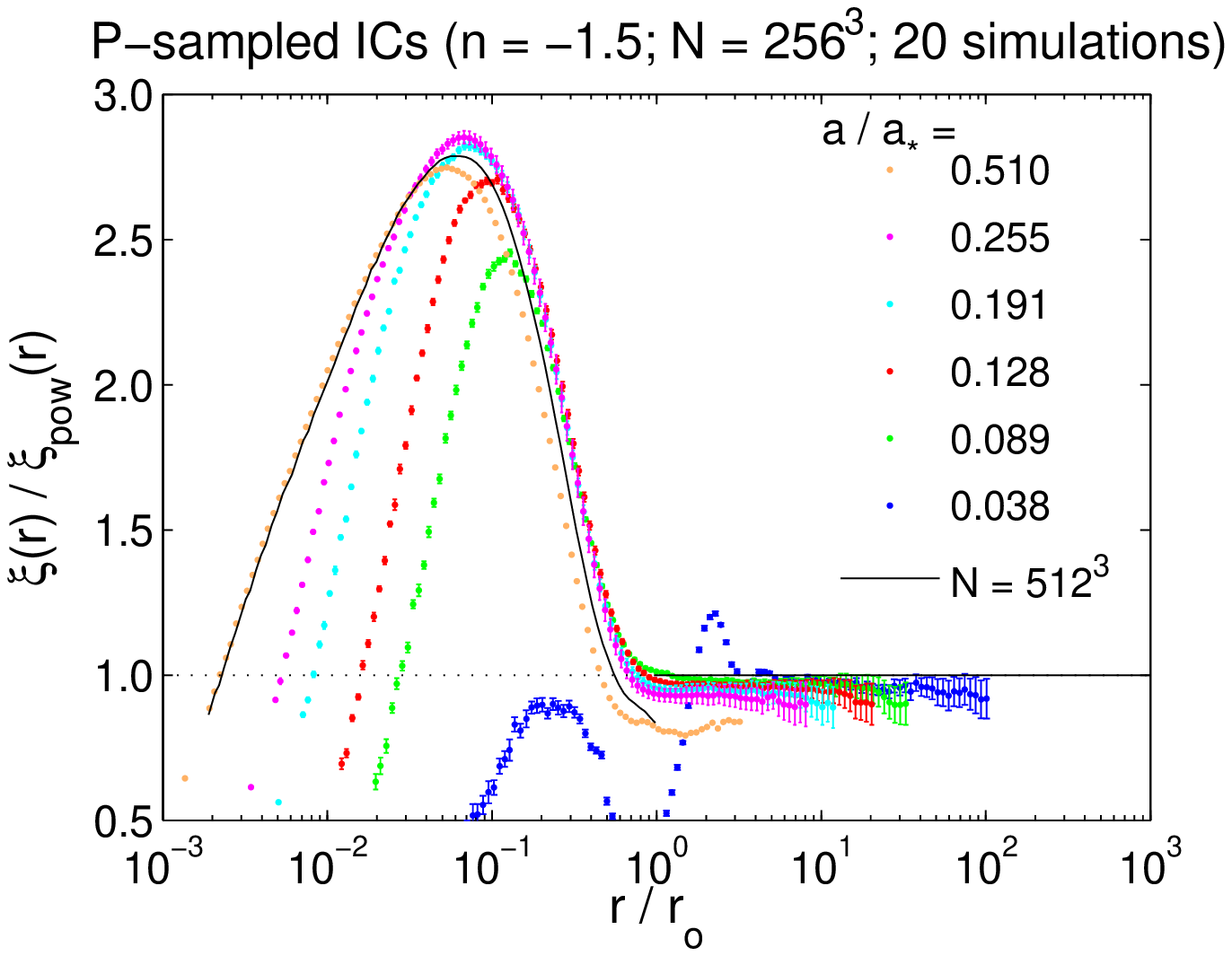, width = 3in}\epsfig{file=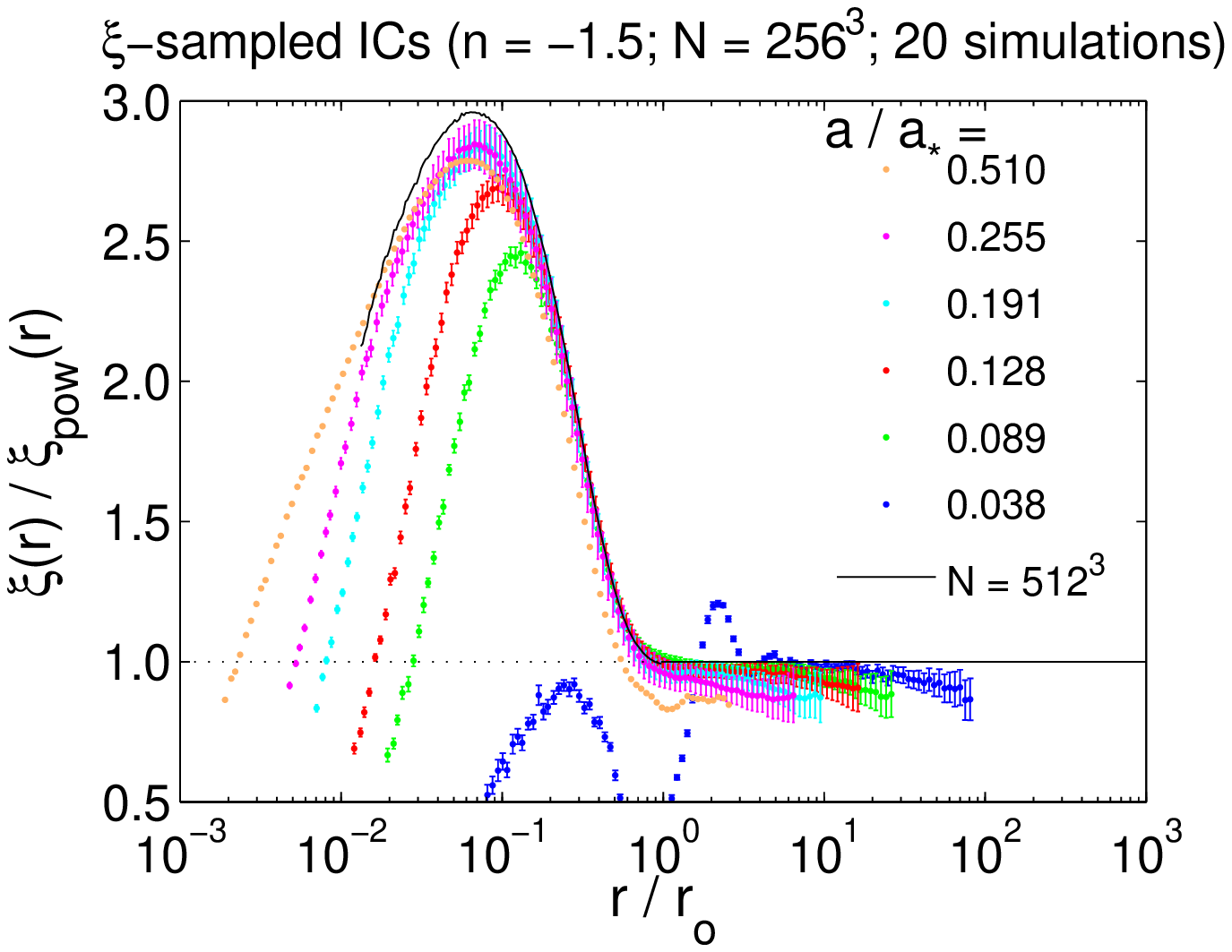, width = 3in}}
\centerline{\epsfig{file=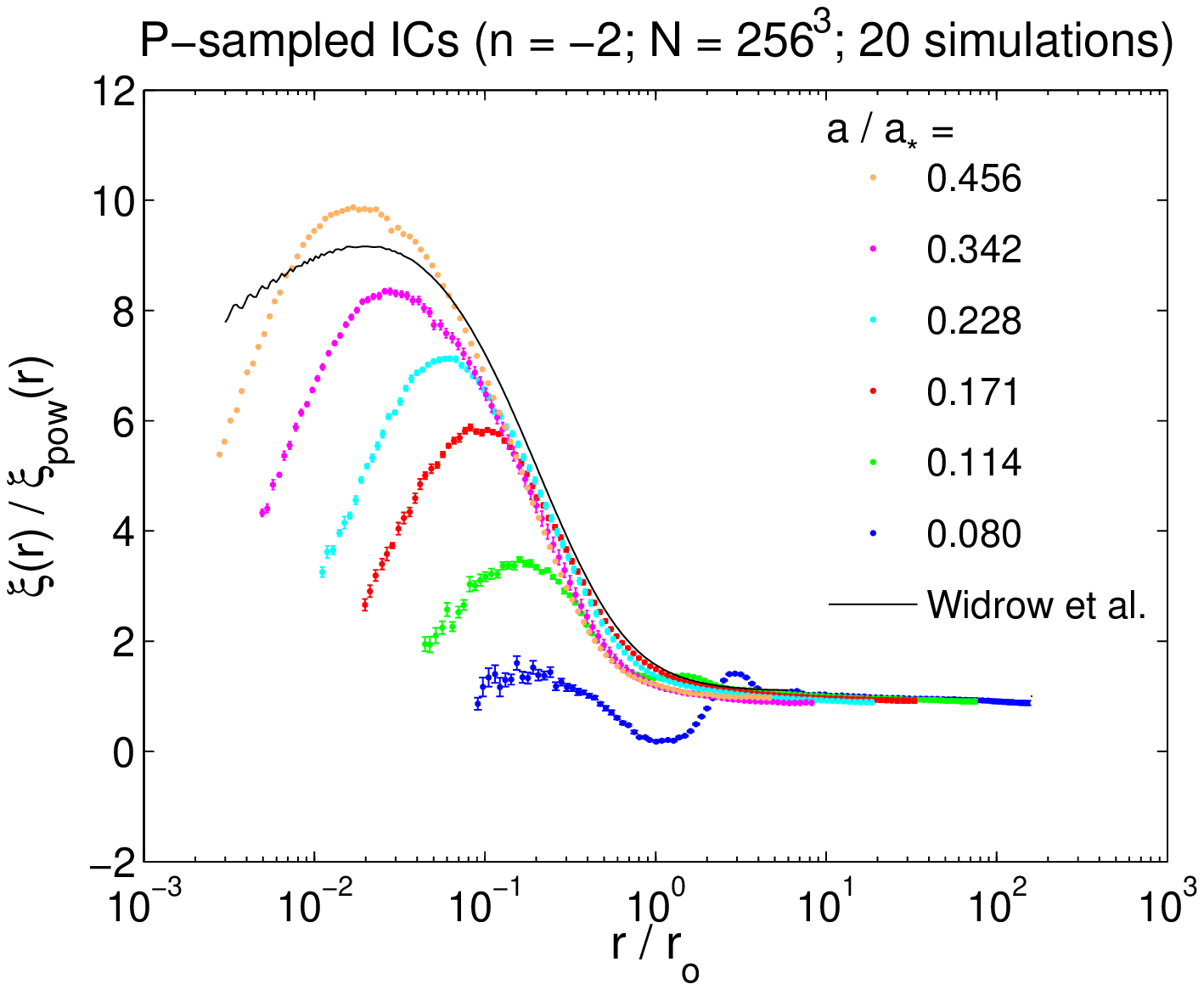, width = 3in}\epsfig{file=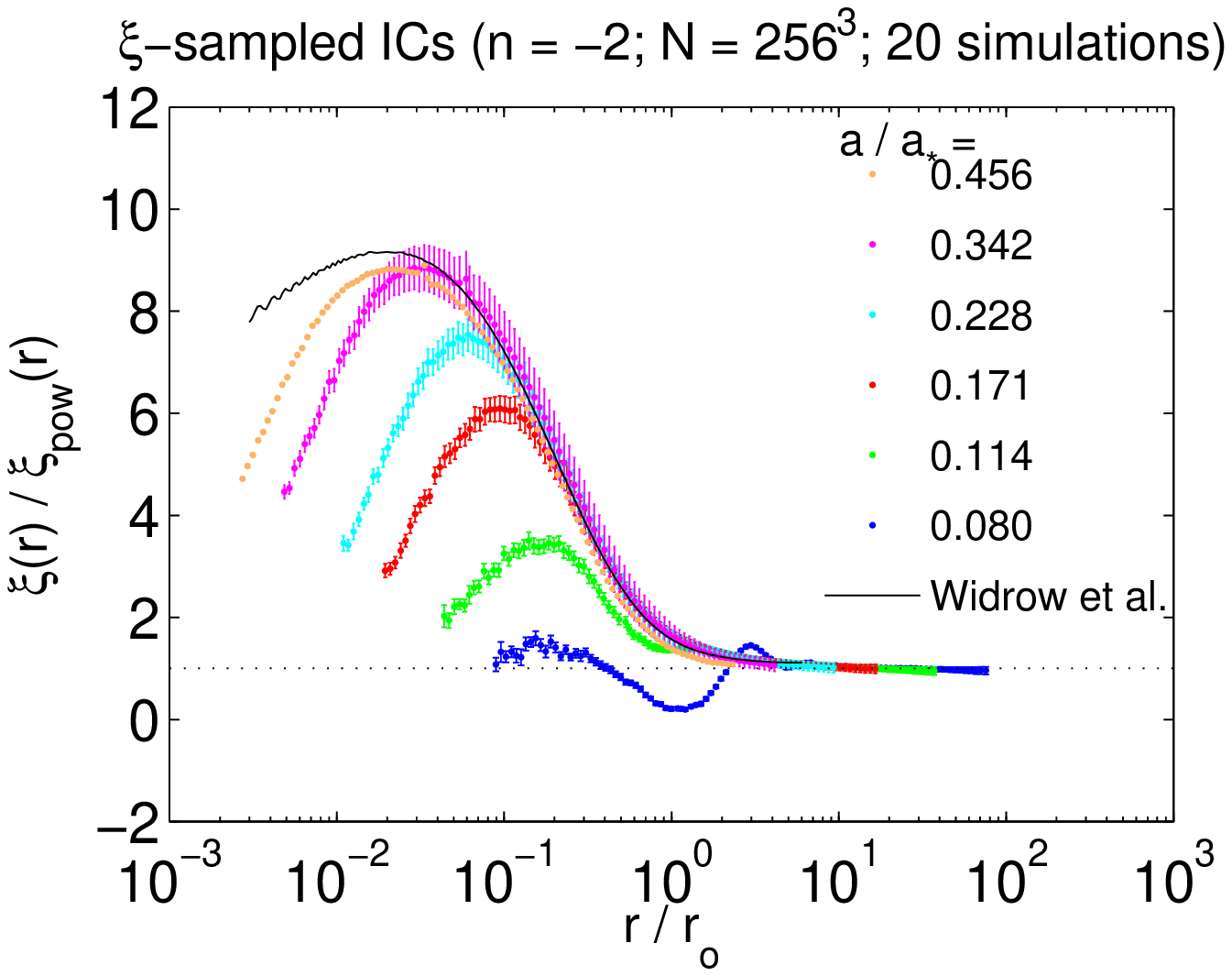, width = 3in}}
\caption{Measured matter autocorrelation functions from conventional 
$P$-sampled (left panels) and $\xi$-sampled (right panels) ensembles of 
simulations. The upper two panels show results from an initially $n = -1$
power spectrum, middle panels show results from $n = -1.5$, and the 
lower two panels show $n = -2$. In each plot the $x$-axis is scaled 
by the non-linear scale, $r_0$, where $\xi_L (r_0) \equiv 1$ so that,
if evolving with the expected self-similar behavior, the outputs should 
lie upon the same locus of points. The $y$-axis is scaled
by $\xi_L (r) = (r_0 / r)^{n+3}$. Black lines show fitting functions for the 
self-similar correlation function from high resolution ($P$-sampled) 
simulations for comparison. Error bars show measured
error on the mean. \emph{Note that the first outputs in each plot are  
affected by transients from initial conditions.}
}\label{fig:xisamp}
\end{figure}

\section{$\xi(r)$ results}
\label{sec:xi}

\subsection{Powerlaw Models}

Fig.~\ref{fig:xisamp} shows my primary results for the self-similar scaling
of the matter correlation function. The x-axis is shown in $r / r_0$ units
where $\xi_L(r_0) \equiv 1$. Insofar as the dark matter clustering 
is negligibly affected by numerical limitations such as the finite scale of 
the box or the scale of the force softening, with this scaling the 
correlation function results from different outputs should all lie upon
the same line. To the extent that this is achieved
the correlation function can be said to evolve with self-similarity and it
is clear from Fig.~\ref{fig:xisamp},  \emph{ excluding the first outputs} which 
are severely affected by transients from initial conditions, that over a wide range of scales
the results from these relatively modest, $N = 256^3$, simulations do fall 
upon the the same locus
as expected. This locus is different for each powerlaw; for steeper power 
spectra (e.g. $n = -2$) power is transferred from large scales to small
scales and the non-linear growth of $\xi(r)$ out paces linear theory
whereas for shallower power spectra (e.g. $n \gtrsim -1$) there is so
much small scale power that the process of halo formation and collapse
causes the non-linear growth to fall behind linear theory in a process
sometimes called ``pre-virialization'' \citep{Davis_Peebles1977}. 
In the language of the halo model \citep{Cooray_Sheth2002} this implies that
the predicted linear theory clustering on small scales is
so high that the amplitude of the 1-halo term is below the linear theory
clustering amplitude on those scales. The $n = -1$ case falls between these
two extremes and the amplitude of the correlation function is both above and
below linear theory, depending on the regime. (For a bracketing case
of an even shallower power spectrum see, e.g., the $n = -0.5$ results in 
\cite[][Appendix A]{Orban_Weinberg2011}.)

\begin{figure}
\centerline{\epsfig{file=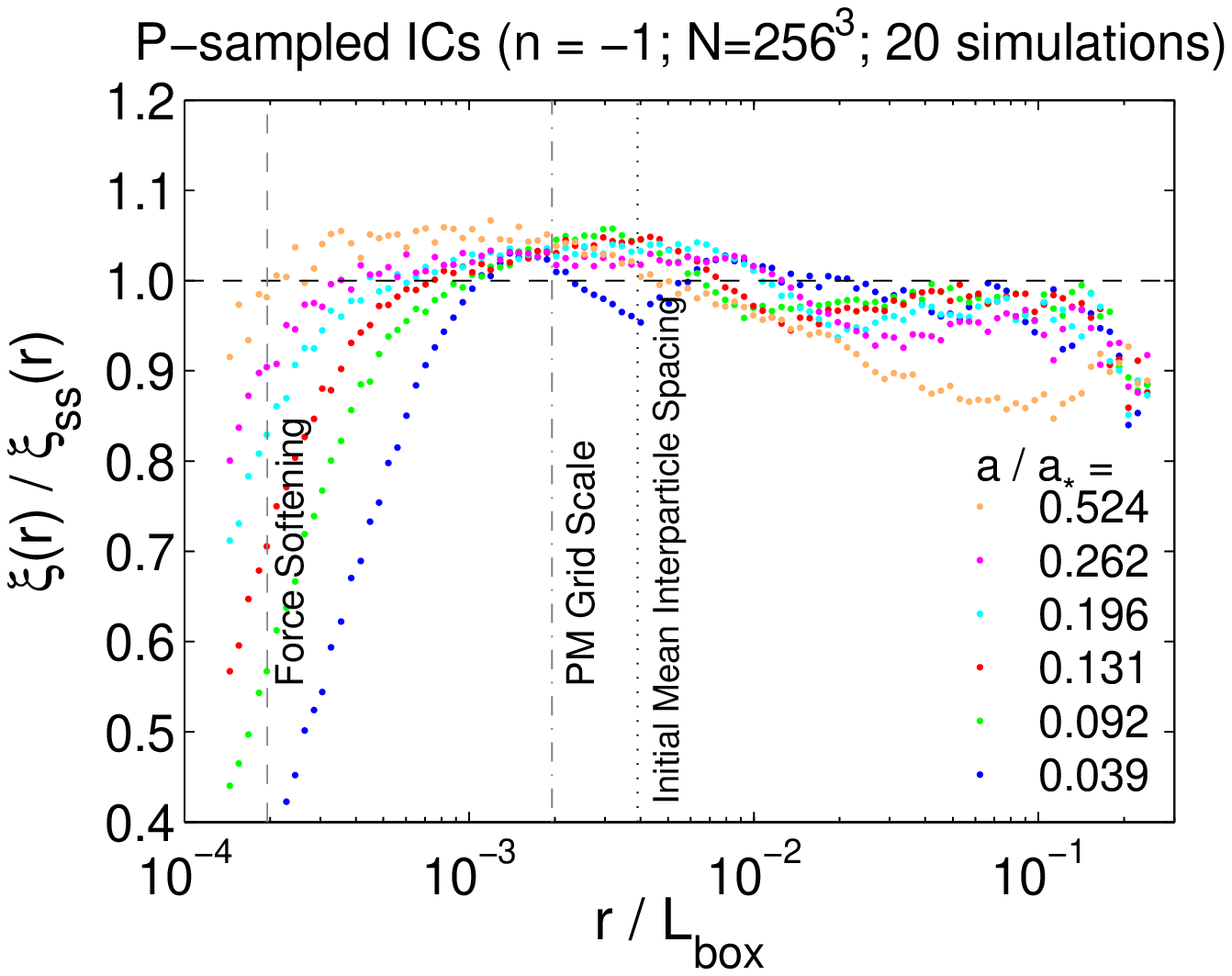, width = 3.15in}\epsfig{file=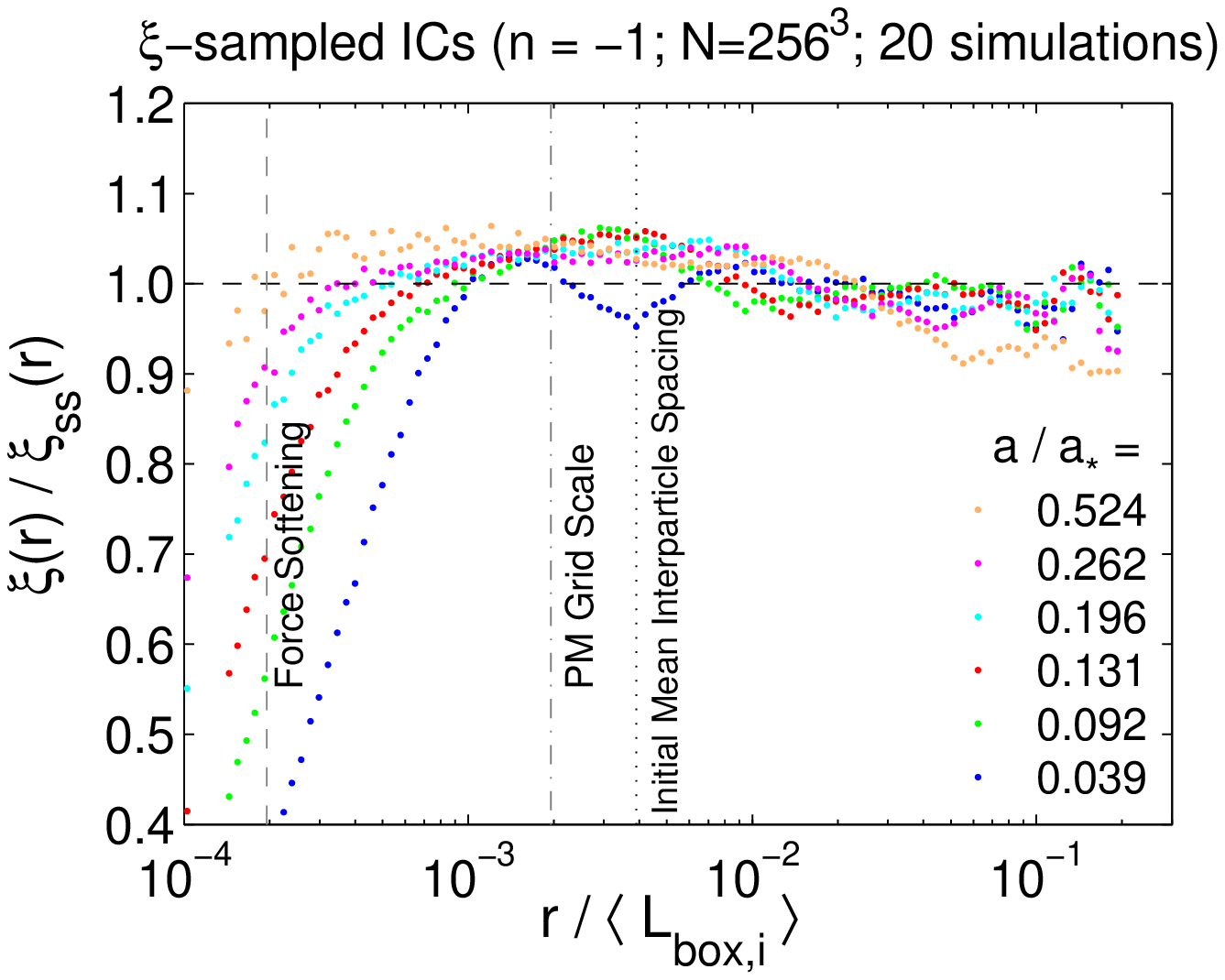, width = 3.15in}}
\centerline{\epsfig{file=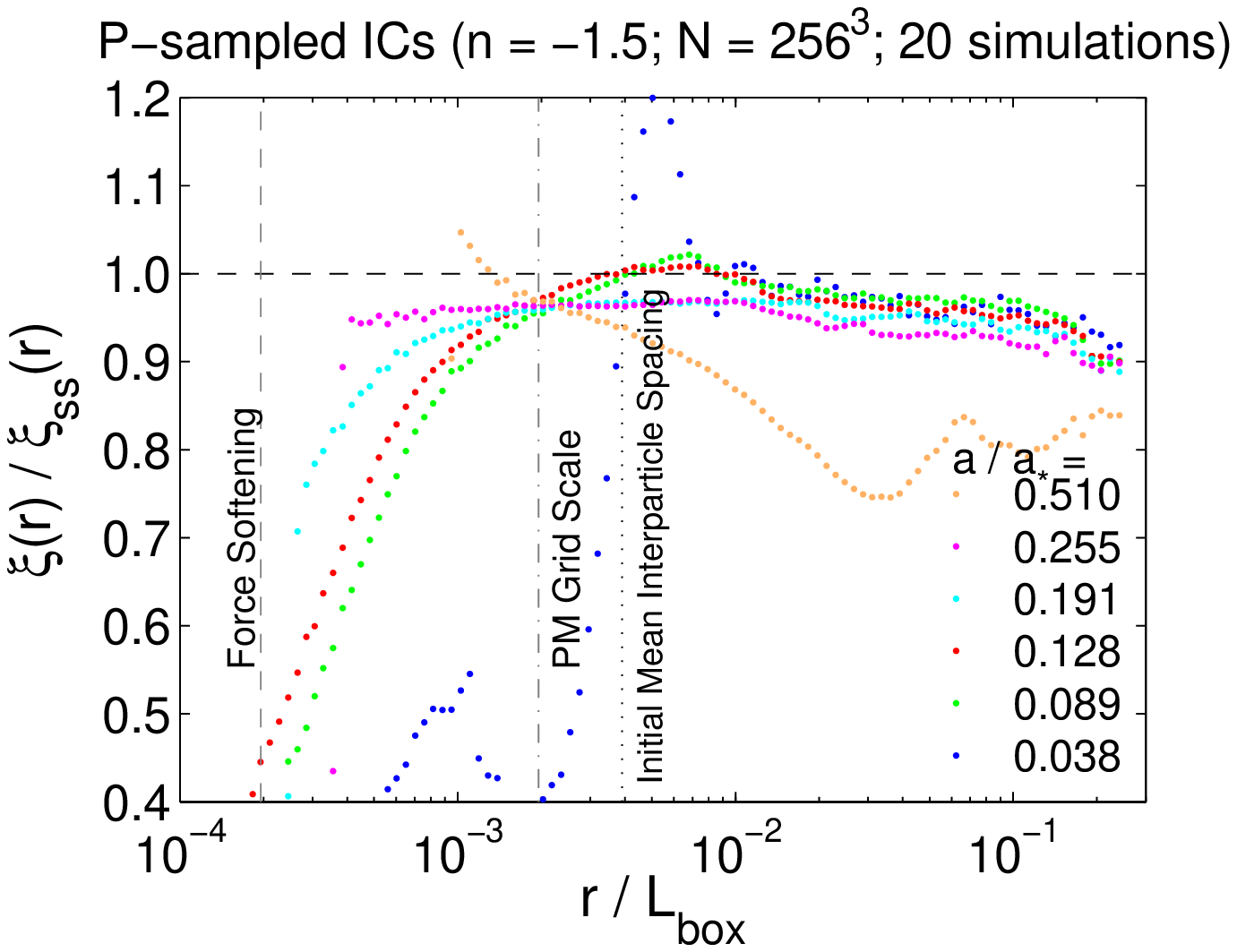, width = 3.15in}\epsfig{file=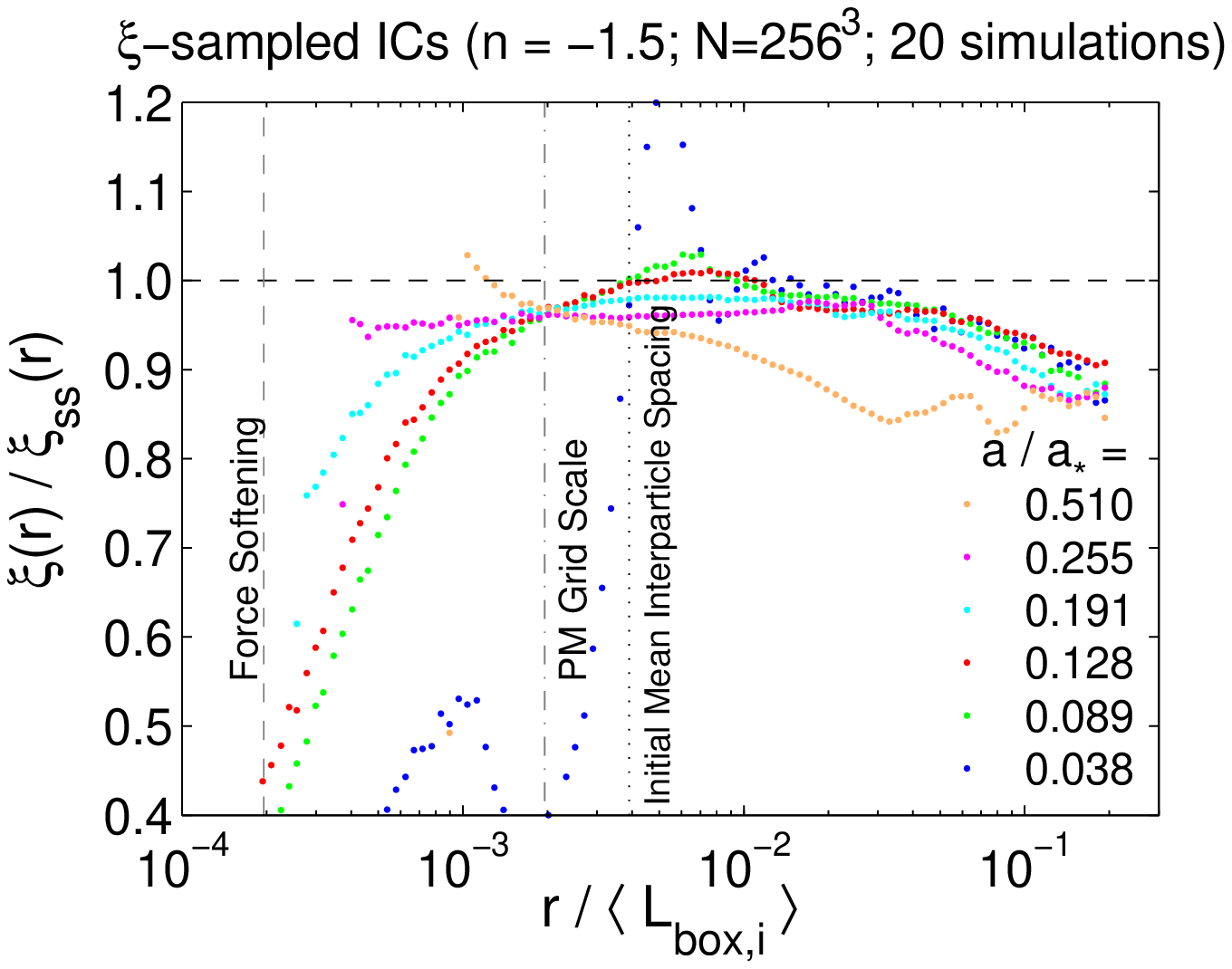, width = 3.15in}}
\centerline{\epsfig{file=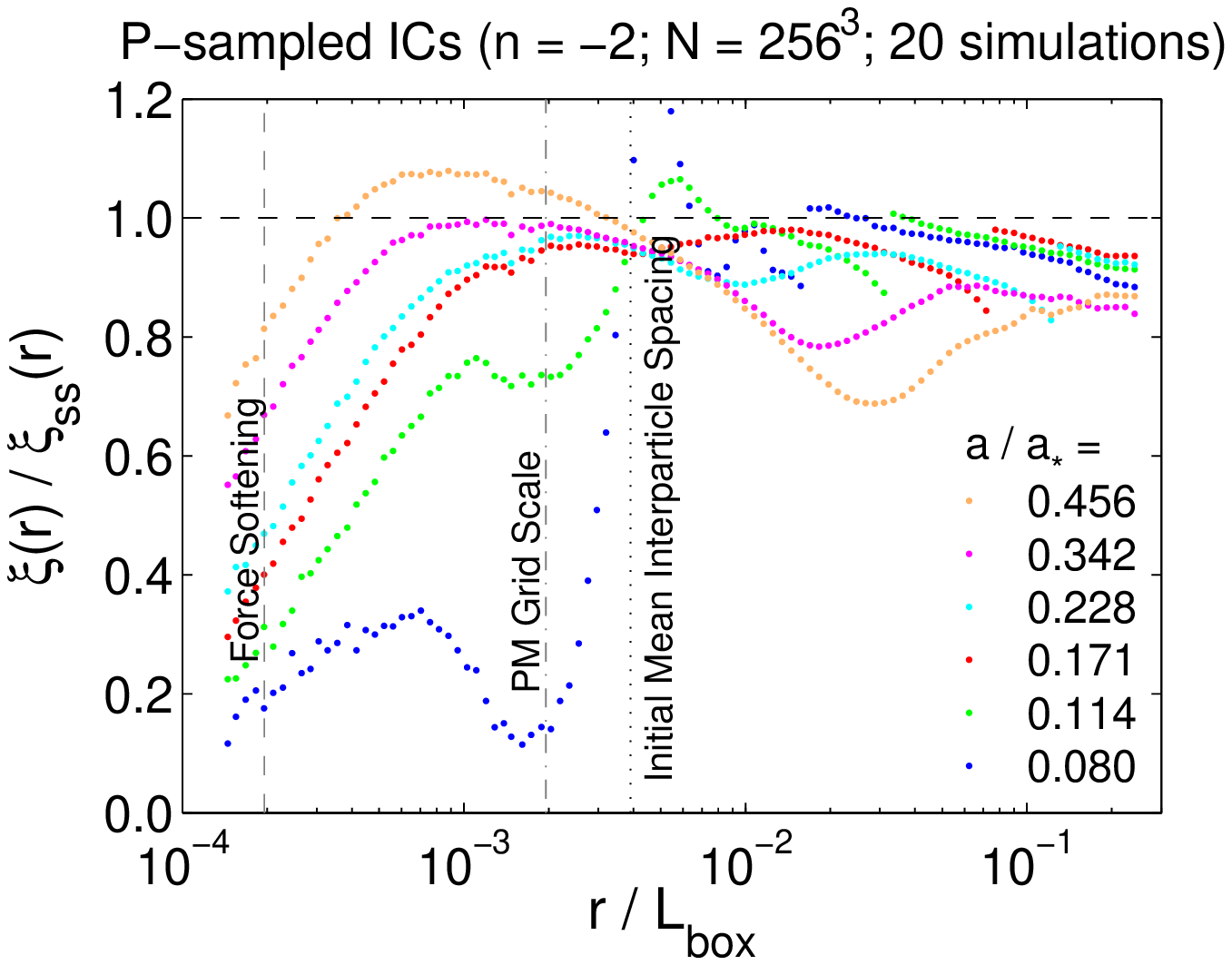, width = 3.15in}\epsfig{file=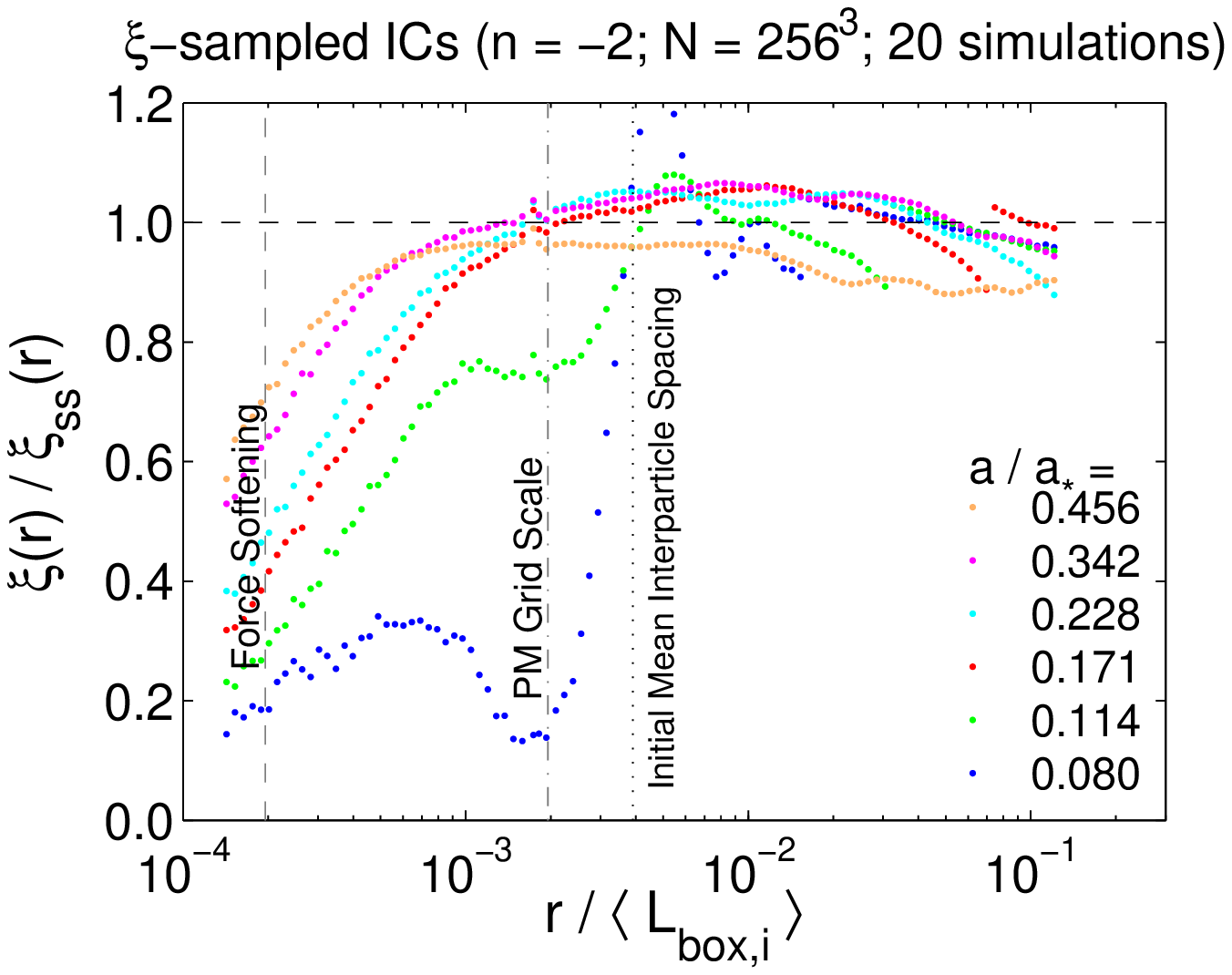, width = 3.15in}}
\caption{Measured correlation functions from simulations (colored points) relative to high-resolution results for the self-similar scaling ($\xi_{ss}(r)$; black lines in Fig.~\ref{fig:xisamp}). Panels are organized as in Fig.~\ref{fig:xisamp} (left panels: $P$-sampled results, right panels: $\xi$-sampled results, $n = -1, -1.5$ and $-2$ from top to bottom). Vertical lines show relevant {\it numerical} scales: the initial mean interparticle spacing (dotted black), the Particle-Mesh Grid Scale (dot-dashed black), and the force softening (dashed black).}\label{fig:xiselfsim}
\end{figure}

In Fig.~\ref{fig:xisamp},  the $\xi$-sampled and $P$-sampled methods generally 
agree well on the shape of the self-similar solution. This is significant for the 
$\xi$-sampled results, on some level verifying the method. Alongside the 
measurements in each case fitting functions for the self-similar correlation
function from higher resolution simulations are shown (black lines). For $n = -1$ and 
$n = -2$ this comparison is made by numerically fourier transforming the 
non-linear power spectrum fitting functions published in 
\cite{Widrow_etal09}; note in the $n = -1$ case I include subtle but important
corrections to their fit at small $k / k_{\rm{nl}}$ as determined in 
\cite[][Appendix A]{Orban_Weinberg2011}.
For $n = -1.5$, I compare with $\xi(r)$ measurements from 10 $P$-sampled
simulations with $N = 512^3$ \citep[][Appendix A]{Orban_Weinberg2011}. 
These high-resolution results are used more quantitatively in 
Fig.~\ref{fig:xiselfsim} where the correlation function results are presented relative to the box size. Overall, the agreement with the high-resolution 
self-similar results is quite good and {\it excluding the initial and final outputs in each case} 
my simulation set tends to match the self-similar evolution to better than 
about 5\% in most outputs and on most scales. This is similar to the 
precision on the results from higher-resolution simulations.
The last output is excluded from this conclusion since the linear theory 
clustering level is so high that one expects departures from the true
non-linear clustering from the suppression of power on the scale of the box.
Also, the correction for the integral constraint, which assumes a
{\it linear theory} correlation function in $\xi_{\rm bias}$ (Eq.~\ref{eq:xibias}), likely 
becomes inaccurate in the highly-clustered regime as well.

Another caveat to the overall good agreement is at small $r / r_o$ especially for early
outputs. Fig.~\ref{fig:xiselfsim} presents the same correlation function measurements in 
Fig.~\ref{fig:xisamp} relative to the scale of the simulation box and shows that 
these deviations from self-similarity are all {\it below} the scale of the initial 
mean interparticle spacing. This is as expected since at best the initial conditions
will only match the self-similar solution down to these separations.
Rather than excluding these scales from Fig.~\ref{fig:xisamp},
they are included to highlight, in Fig.~\ref{fig:xiselfsim}, that 
as structure evolves the self-similar behavior extends further and further below this
scale, in some cases approaching the force softening. This result is non-trivial and
difficult to anticipate from first principles. 

It bears mentioning some of the previous work on how non-linear clustering proceeds near or below 
the scale of the initial mean interparticle spacing. \cite{Little_etal1991},
using $n = -1$ simulations, show that Fourier modes in the non-linear 
regime are largely determined by the collapse of large-scale modes rather than
by evolution of power initially on those scales. This nicely explains the 
trend in Fig.~\ref{fig:xiselfsim} for later outputs to match better with the self-similar 
solution on small scales and why the poisson noise in the dark matter density on those
small scales does not prevent this from happening. However, Joyce et al. \cite{Joyce_etal2009}
and collaborators have argued that the common practice of setting the force
softening significantly smaller than the initial mean interparticle spacing
(as in the simulations presented here) introduces errors which arise from the 
possibility that with this choice
the equations of motion for the particles are no longer true to the
Vlasov-Poisson fluid equations. Despite this, their results concur with 
Fig.~\ref{fig:xiselfsim} that $\xi(r)$ can reliably be modeled below the scale of the mean interparticle
spacing. According to \cite{Joyce_etal2009} the main effect of aggressive 
force softening is to cause $\sim 5\%$
disagreement with the true non-linear $\xi(r)$ on scales 
{\it larger} than the mean interparticle spacing. The $P$-sampled results
shown in Fig.~\ref{fig:xiselfsim} are in qualitative agreement with the 
simulations presented in \cite{Joyce_etal2009} in the sense that accurate
non-linear behavior is observed below the mean interparticle spacing
and on larger scales the measurements are consistent with the self-similar
solution also at the level of $\sim 5\%$. Although beyond the scope of this
paper, it would be interesting to run the $P$-sampled simulation set with less
aggressive force softening (e.g. half the mean interparticle spacing) to test
if the measured error on the mean $\xi(r)$  
is detectably smaller, as predicted in \cite{Joyce_etal2009}. At any rate,
for all three powerlaws the self-similar behavior extends well below the 
scale of the mean interparticle spacing; it does not significantly
depend on whether power is being rapidly ``transferred'' to smaller scales as 
for $n = -1.5$ and $n = -2$ or whether the non-linear growth proceeds less
quickly than the linear theory prediction on small scales 
(i.e. $r < r_{0}$), as for $n = -1$.

\subsection{Powerlaw Times a Bump Results}

\begin{figure}
\centerline{\epsfig{file=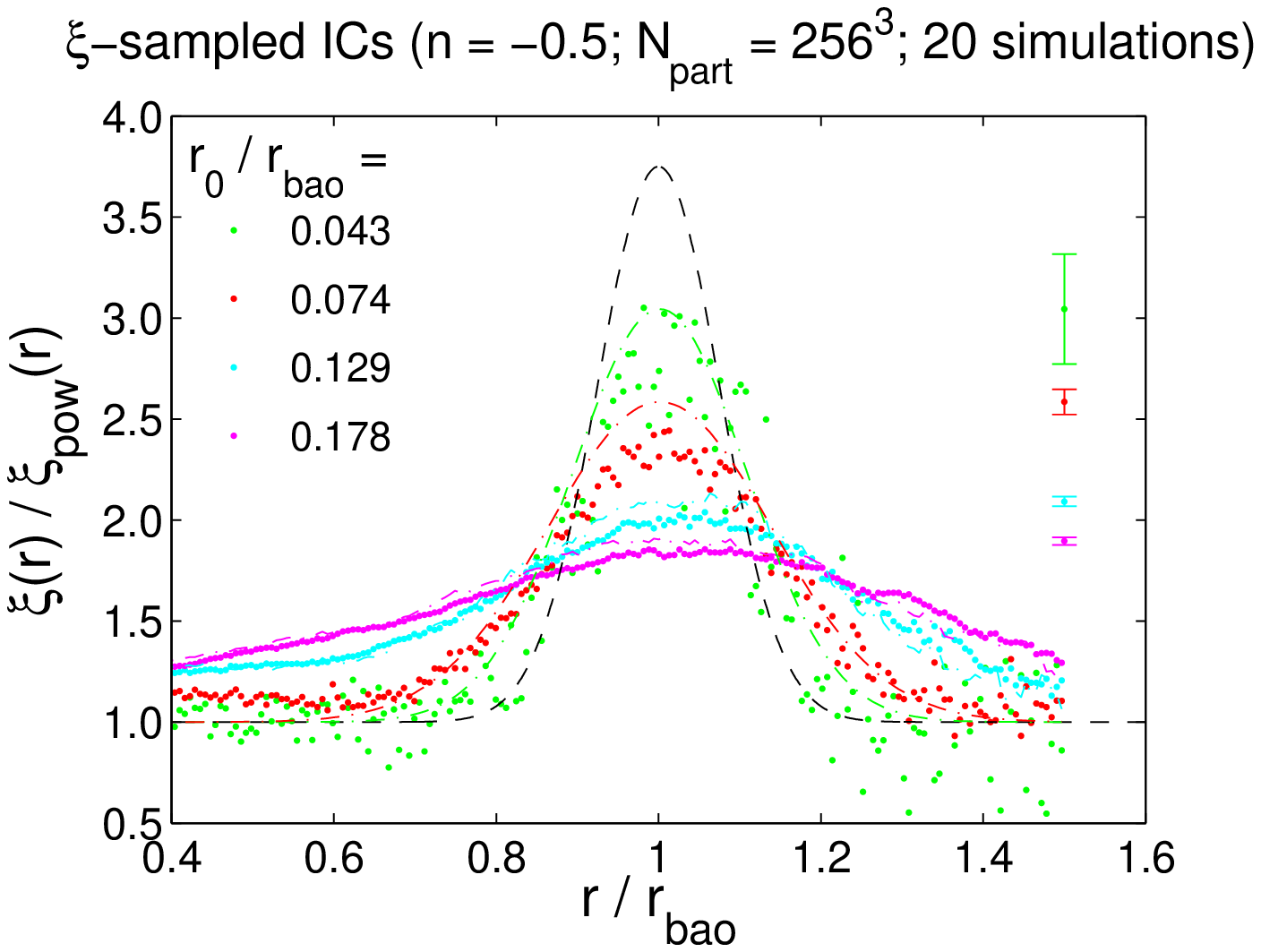, width = 3.15in}\epsfig{file=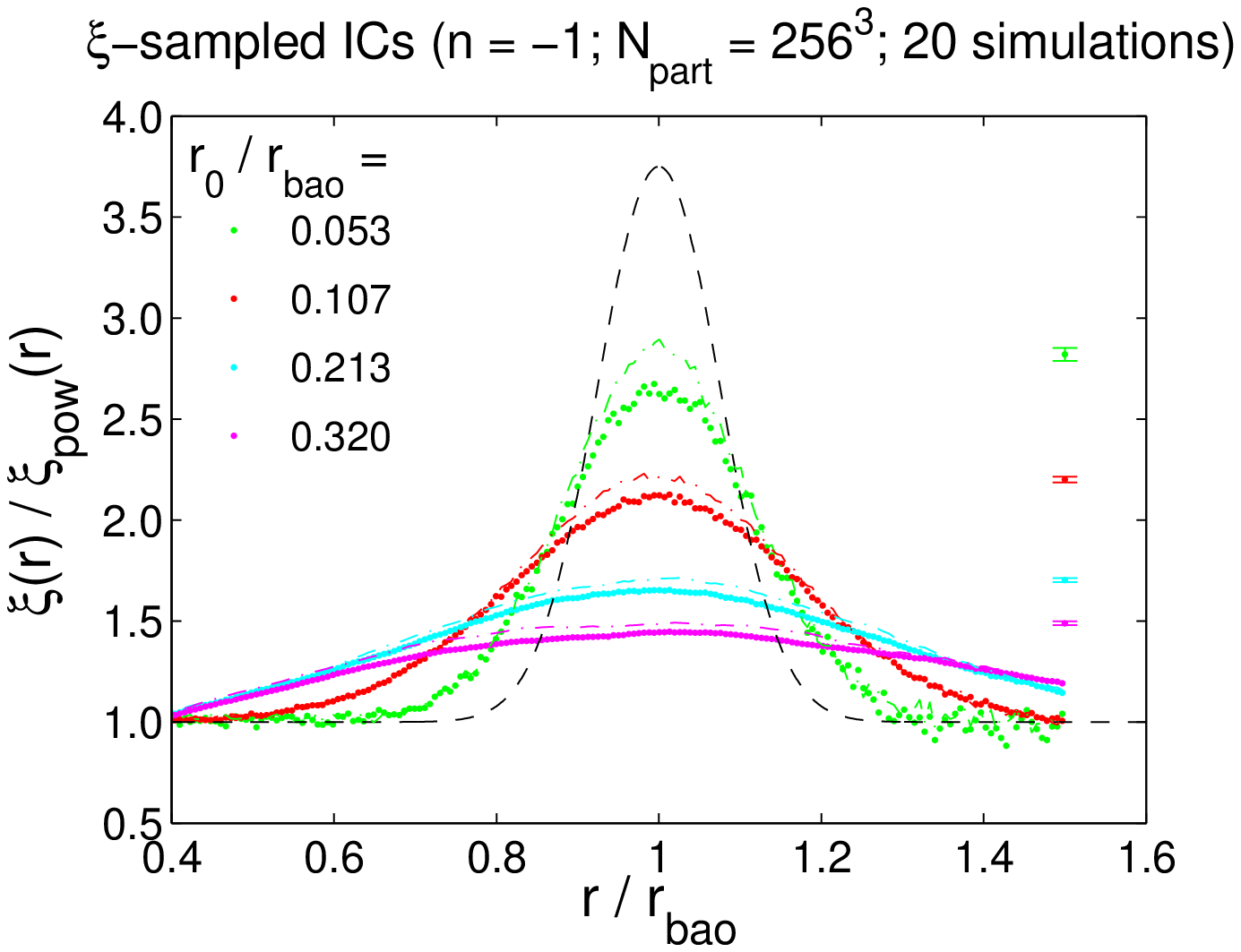, width = 3.15in}}
\centerline{\epsfig{file=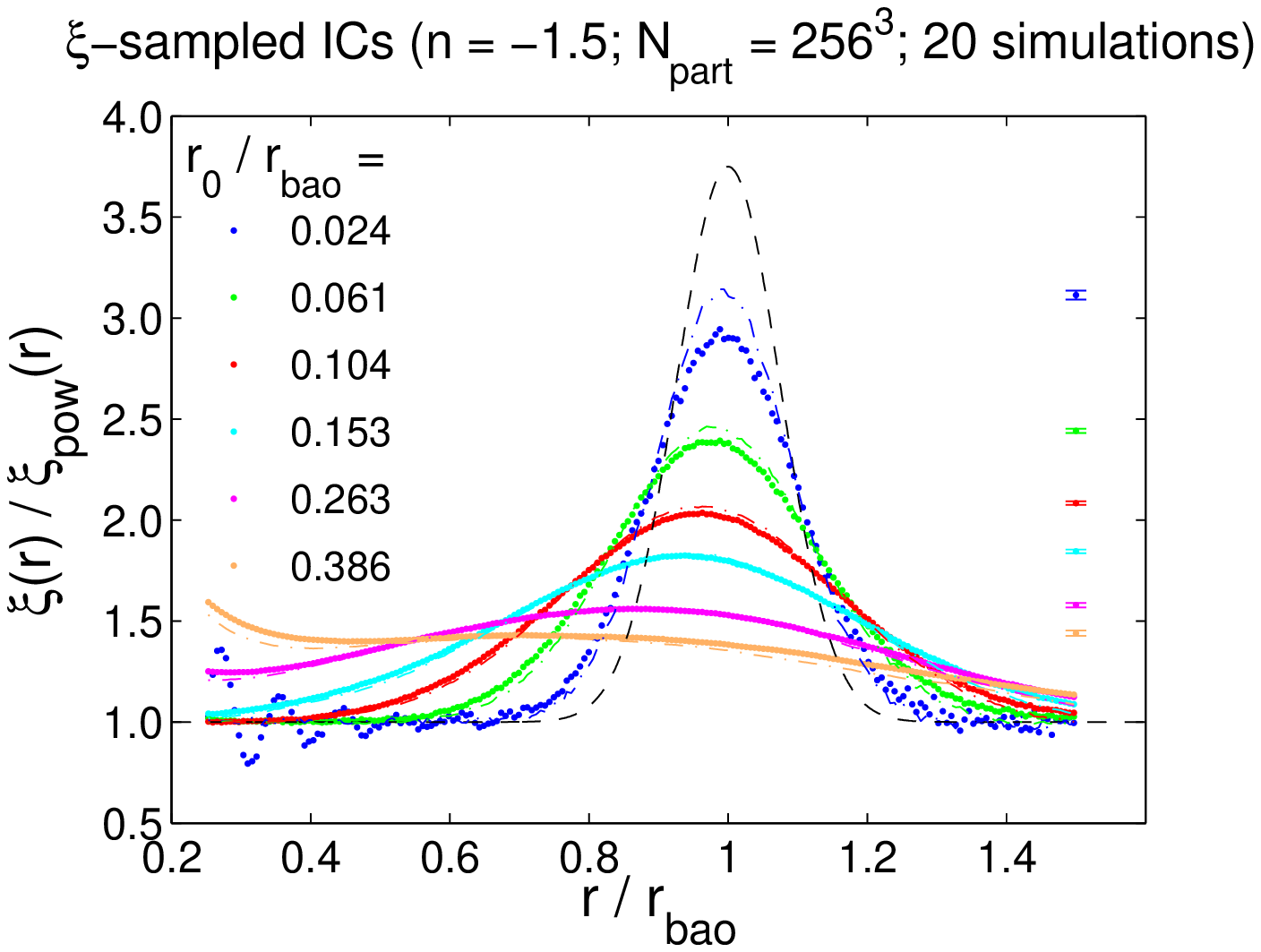, width = 3.15in}}
\vspace{-0.1in}
\caption{Correlation function results from ensembles of 20 $\xi$-sampled simulations using initial conditions consistent with a powerlaw times a gaussian bump as a simplified model of baryon acoustic oscillations. Dot-dashed lines show results from the high-resolution simulations presented in \cite[][Fig. 3]{Orban_Weinberg2011}. Typical errors on the mean for the $\xi$-sampled results are shown offset to the right. The initial bump width and height from the initial conditions is shown with a dashed black line. Note that in the $n = -0.5$ panel in the top left, for ease of comparison the dot-dashed lines are derived from gaussian fits to the $P$-sampled results instead of simply presenting the actual correlation function measurement as in the other panels because these measurements are somewhat noisy.
}\label{fig:powbump}
\end{figure}

\begin{figure}
\centerline{\epsfig{file=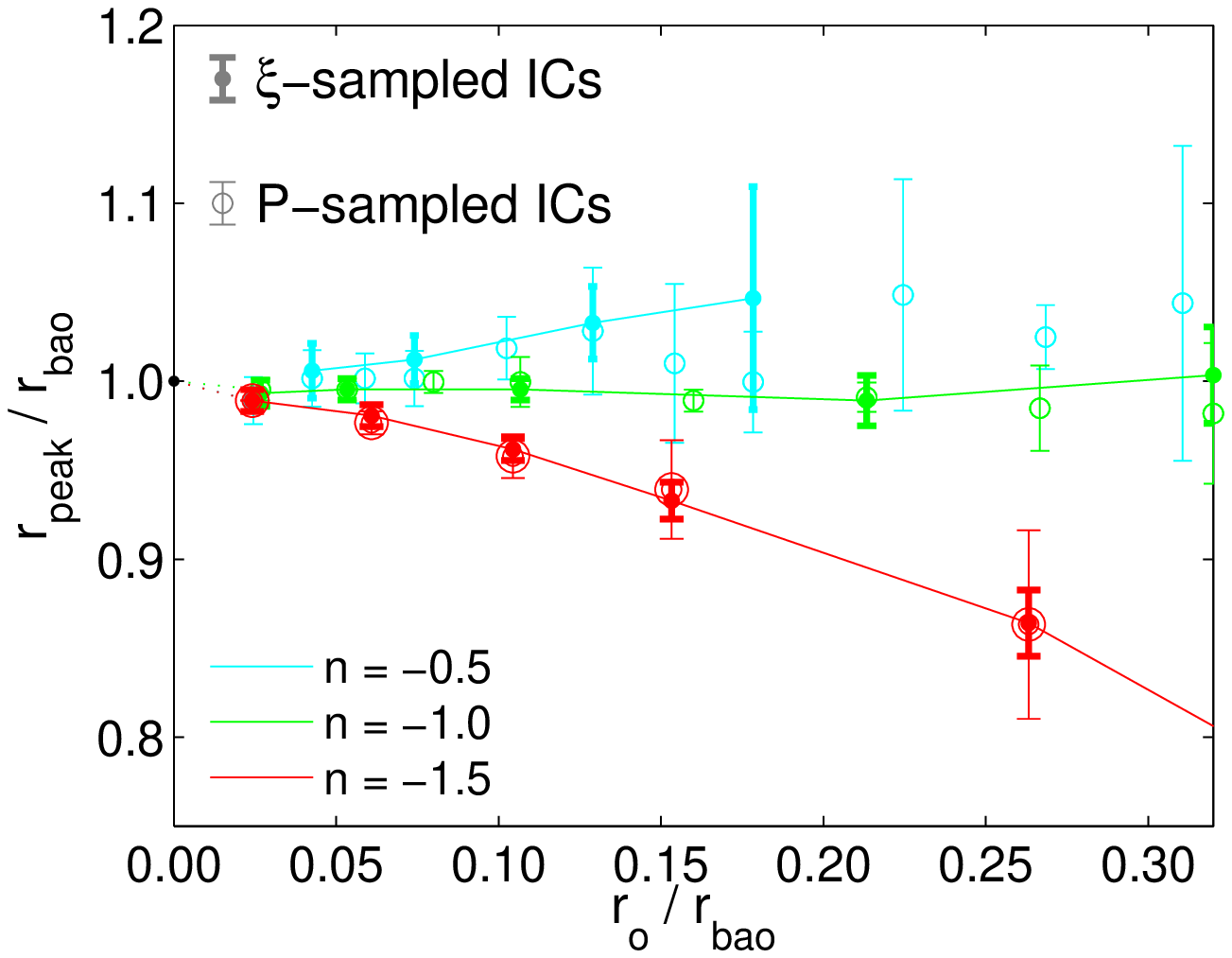, width = 3.15in}\epsfig{file=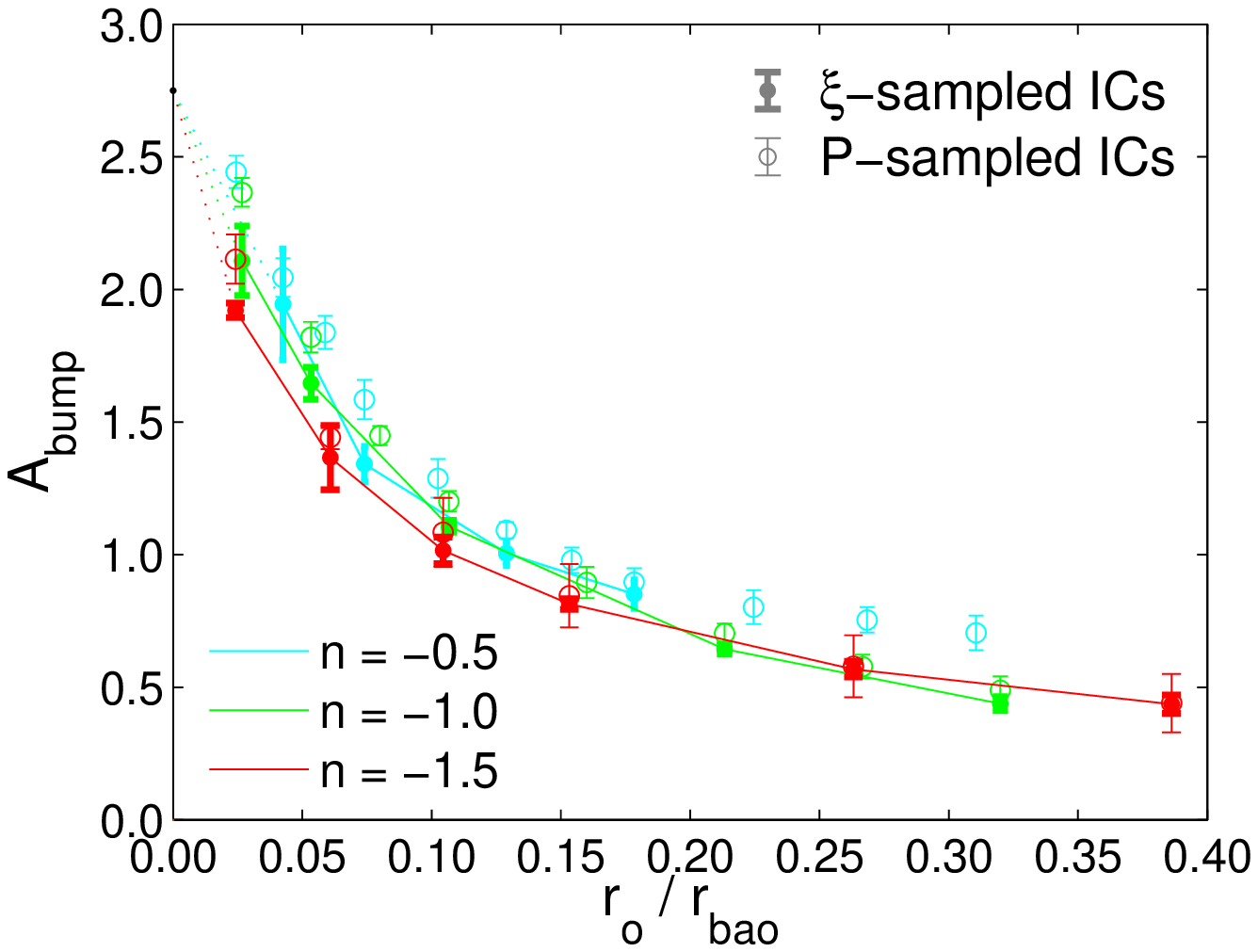, width = 3.15in}}
\centerline{\epsfig{file=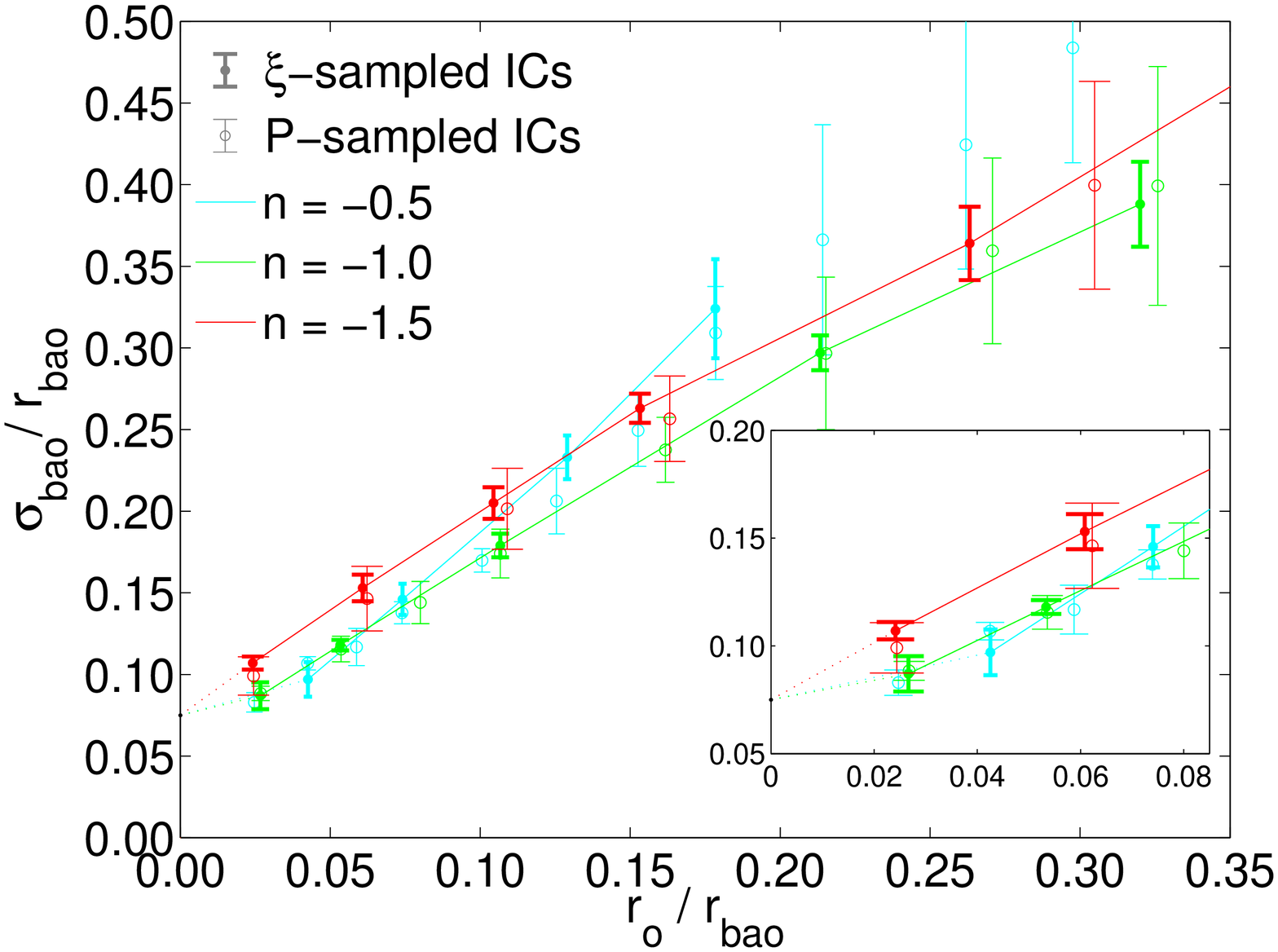, width = 3.15in}\epsfig{file=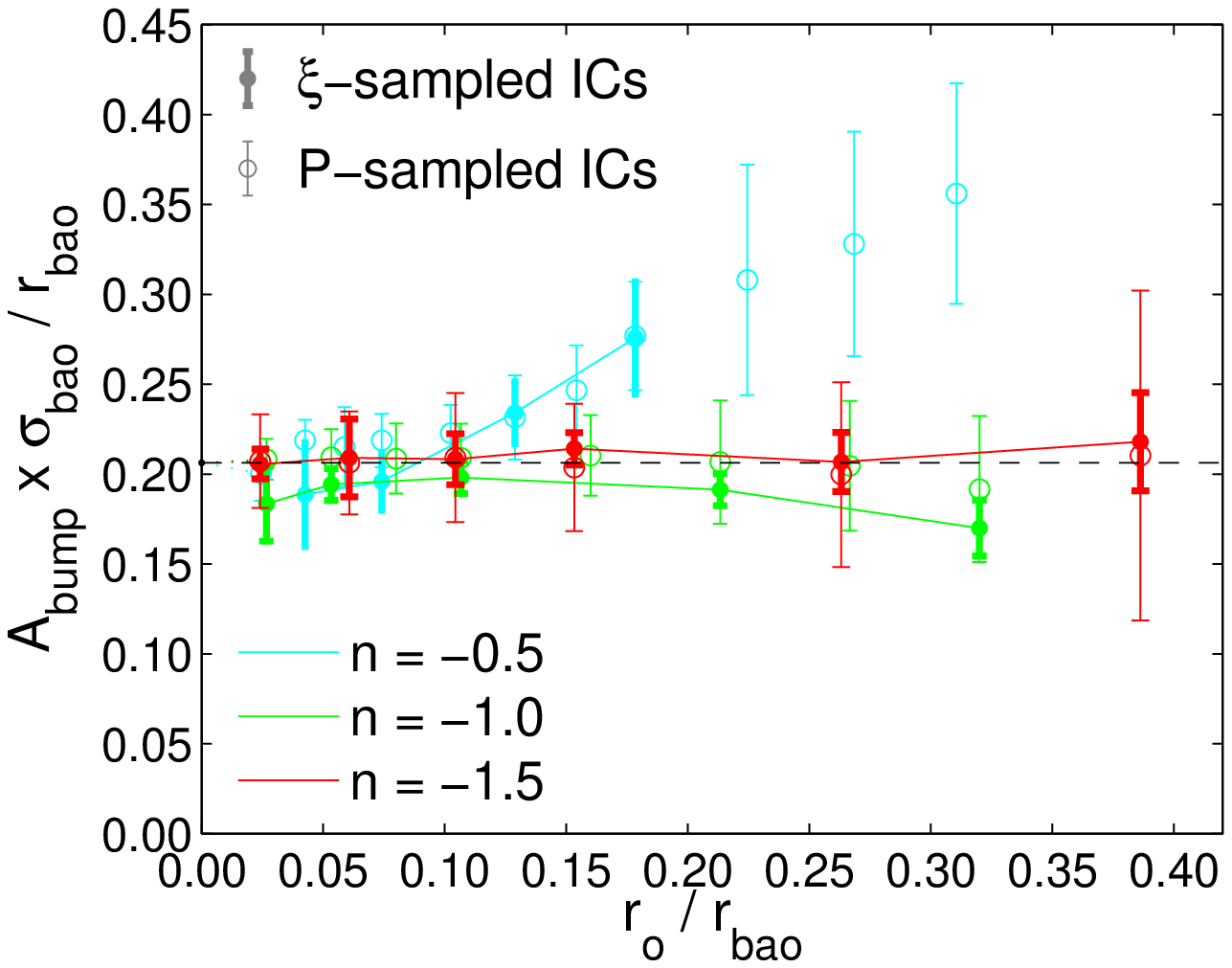, width = 3.15in}}
\vspace{-0.1in}
\caption{ Results from gaussian fits to the simulation results presented in Fig.~\ref{fig:powbump}.  Each plot shows best-fit quantities versus $r_o / r_{\rm bao}$ (i.e. the time variable) for all three powerlaws, cyan for $n = -0.5$, green for $n = -1$ and red for $n = -1.5$. Upper left: the best-fit position of the peak. Upper right: the best fit amplitude of the BAO feature. Bottom left: gaussian width of the BAO feature.  Bottom right: the normalized area of the BAO feature. Errorbars throughout are derived from jackknife error estimation.
}\label{fig:baoquant}
\end{figure}

As discussed in depth in \cite{Orban_Weinberg2011}, a real-space powerlaw times a bump
can be used as a self-similar numerical test in addition to providing 
insight into the non-linear physics of the evolution of the BAO feature.
In this case, 
\begin{equation}
  \xi_{L}(r) = \left(\frac{r_0}{r} \right)^{n+3} (1 + A_{\rm{bump}} \, e^{-(r-r_{\rm{bao}})^2/2\sigma_{\rm{bao}}^2}),
  \label{eq:powgaus}
\end{equation}
and for resemblance to the $\Lambda$CDM correlation function 
I chose $A_{\rm{bump}} = 2.75$, $\sigma_{\rm{bao}}/r_{\rm{bao}} = 0.075$, 
and powerlaws of $n = -0.5$, $-1$, and $-1.5$. Unlike $\Lambda$CDM,
this setup can be evolved much further than $\sigma_8 \sim 1$ to 
investigate the non-linear physics of the problem. For each powerlaw, 
in Fig.~\ref{fig:powbump} I compare results from 20 $\xi$-sampled simulations
with $N = 256^3$, $r_{\rm{bao}} / \langle L_{{\rm box},i} \rangle = 1/20$ to the results of 7 
$P$-sampled, $N = 512^3$, $r_{\rm{bao}} / L_{\rm{box}} = 1/20$ 
simulations from \cite{Orban_Weinberg2011}. In Fig.~\ref{fig:powbump} these 
$P$-sampled results are shown with dot-dashed lines of various colors 
corresponding to different outputs. Since the first two outputs from the $P$-sampled
$n = -0.5$ simulations are noisy because of the very low clustering amplitude, 
Fig.~\ref{fig:powbump} presents the best fit gaussians to those results for ease of comparison.
All other dot-dashed lines are the mean correlation function results from the 
$P$-sampled simulations.
Error bars in Fig.~\ref{fig:powbump} show the error on the mean for the $\xi$-sampled
results. Qualitatively, the correlation function results agree well and importantly the non-linear
shift in the $n = -1.5$ results and lack of shift in the $n = -0.5$ and $-1$ results
are consistent. This conclusion should be reassuring to the wider effort
to characterize the non-linear shift of the BAO peak using standard 
$P$-sampled simulations. 

A quantitative comparison of the results in Fig.~\ref{fig:powbump} is presented in 
Fig.~\ref{fig:baoquant}. Here the $\xi$-sampled results are shown with solid 
points with thick error bars, and the $P$-sampled results are shown as 
open circles with thin error bars, both with colors corresponding to the powerlaw
(cyan for $n = -0.5$, green for $n = -1$, and red for $n = -1.5$).
As in Fig.~5 of Orban \& Weinberg \cite{Orban_Weinberg2011}, the error bars
for both methods come from jackknife error estimation by sequentially omitting one of
the realizations and computing the best fit gaussian and shift of the peak. 
The $\xi$-sampled results typically have tighter error bars than 
the $P$-sampled results because more $\xi$-sampled simulations were performed.

The upper left panel of Fig.~\ref{fig:baoquant} echoes what was said earlier 
that the non-linear shift of the BAO peak is consistent between the two 
methods. The $n = -0.5$ results for both methods show some preference for a BAO 
shift to slightly {\it larger} scales, however the shift in this case is degenerate with 
the broadening (notice that the error bars in the bottom left plot are relatively large at later outputs) 
and the error bars are consistent with no movement 
of the BAO peak. A real movement of the peak to larger scales would have been counter-intuitive since the
non-linear physics of the shift stems from a small but non-negligible 
tendency for particle pairs with initial separations of $r = r_{\rm bao}$ 
to be found in regions with a slight overdensity, causing (on average) a very small movement 
{\it inward} \cite{Eisenstein_etal2007}.

Fortunately for BAO surveys, the broadening of the BAO feature from the growth of structure
 is a much larger effect than the non-linear shift. The bottom left panel of Fig.~\ref{fig:baoquant} highlights
the results for broadening of the gaussian width of the BAO feature as it evolves from its 
initial value of $\sigma_{\rm bao} / r_{\rm bao} = 0.075$. 
The results from the two methods compare well and there are no pairs of points
from any particular output or powerlaw that are statistically inconsistent 
with each other. In Fig.~5 of Orban \& Weinberg \cite{Orban_Weinberg2011} 
a very similar plot compared these $P$-sampled results to a simple diffusion model
inspired by \cite{Eisenstein_etal2007}. This earlier comparison was 
reasonably good and remarkably the $n = -1.5$ results agreed with an {\it ab initio}
prediction of the diffusion model. The agreement between the $\xi$-sampled
and $P$-sampled results argues that this same physics is properly included in 
$\xi$-sampled simulations and, e.g., that including the DC mode fluctuations 
in overdensity does little to change this result.

The top right panel of Fig.~\ref{fig:baoquant} presents the results for the 
amplitude of the BAO feature. At later outputs the two methods agree
well, however there is some tension with the first few outputs.
This would be concerning except that the $P$-sampled $N = 256^3$ results in 
Fig.~8 of Orban \& Weinberg \cite{Orban_Weinberg2011}
show a similar decrement in bump amplitude compared to $N = 512^3$ $P$-sampled 
simulations at early outputs. This mismatch seems to be some finite-particle numerical effect
as opposed to some orthogonal concern relating to 
box scale cutoffs of large scale power. The bottom right panel shows the results for the 
normalized area of the bump, $ A_{\rm bump} \times \sigma_{\rm bao} / r_{\rm bao}$
which tends to be constant in spite of the non-linear evolution of 
the BAO feature in agreement with the diffusion model discussed in Orban \& Weinberg \cite{Orban_Weinberg2011}. 
In the $\xi$-sampled simulations at early outputs 
the bump area falls somewhat below its initial value for both $n = -0.5$ and $-1$ by about one sigma. 
This can be attributed to the decrement of $A_{\rm bump}$ since $\sigma_{\rm bao}$ evolves as expected, but
more importantly this tension with the constant-bump-area evolution at these early outputs 
seems to corroborate the conclusion that it is merely an inaccuracy from using $N = 256^3$ particles instead of $N = 512^3$.

\section{Box-to-Box Variance of the Correlation Function}
\label{sec:boxtobox}

\subsection{Preliminaries}
\label{sec:prelim}

Having explored the ensemble-averaged predictions for the mean $\xi(r)$,
in this section I compare the results for the box-to-box variance of $\xi(r)$
from the $\xi$-sampled and $P$-sampled methods, focusing on separations
approaching the box scale ($r \gtrsim L_{\rm box} / 10$). 
While the variance (or, more generally, 
covariance) of statistics like $\xi(r)$ is important for surveys so as to 
precisely and accurately infer cosmological constraints from a finite data set 
\cite{Habib_etal2007,Meiksin_White1999,Scoccimarro_etal1999,Cohn2006,Hamilton_etal2006,Takahashi_etal2009},  
the primary goal of this section is somewhat more prosaic. Namely,
if the box-to-box variance of $\xi(r)$
from one or the other method is substantially larger then substantially more simulations must 
be performed via this method to obtain the same precision on the mean $\xi(r)$.
This would be the only reason to perform additional simulations since 
\S~\ref{sec:intbiassec} and \S~\ref{sec:xi} show
that as long as the integral constraint correction is applied to $P$-sampled
correlation function measurements, the mean $\xi(r)$ is consistent between
the two methods.

The box-to-box variance comes from the (usual) definition,
\begin{equation}
{\rm Var} ( \xi ) = \langle (\xi_{{\rm uni},i}(r) - \xi(r))^2 \rangle = \frac{1}{N_{\rm sims}-1} \sum_{i=1}^{N_{\rm sims}} \, \left( \xi_{{\rm uni},i}(r) - \xi(r) \right)^2   \label{eq:varxi}
\end{equation}
where $\xi_{{\rm uni},i}(r)$ is a correlation function measurement from an individual box.
 For $\xi$-sampled simulations, since the overdensity of each box is perturbed from
the mean density of the true cosmology, one must use Eq.~\ref{eq:xiunii3} to ``convert'' $\xi_{{\rm box},i}(r)$ 
(a statistic that assumes incorrectly that the overdensity of the box is zero) 
to $\xi_{{\rm uni},i}(r)$ as discussed in \S~\ref{sec:cosmostats}. For $P$-sampled simulations this step is 
unnecessary because every realization has zero overdensity by design 
and consequently $\xi_{{\rm uni},i}(r) = \xi_{{\rm box},i}(r)$. 

This section will discuss three different sources of box-to-box variance (or, equivalently, 
three different considerations that degrade the precision on the mean $\xi(r)$ from 
a finite number of simulations). 
These sources are: (1) variance from ignorance of the overdensity of the realizations, 
(2) variance from measuring $\xi(r)$ from a finite number of randomly realized density fields, and
(3) variance from correlations on weakly to strongly non-linear scales. 

The first item is mentioned only for completeness. As discussed in \S~\ref{sec:cosmostats}, the 
$\xi$-sampled correlation function estimator in Eq.~\ref{eq:xiunii3} is implicitly ``informed'' 
of the overdensity and likewise the $P$-sampled estimator is informed of the 
overdensity in the sense that the overdensity of each box is identically zero. 
Were this {\it not} the case, then on large scales where the correlation is weak, 
following the discussion in \S~\ref{sec:better} the correlation function measurement would yield the overdensity of the box
at that epoch, $\xi_{{\rm uni},i}(r) \approx 2 \Delta_i$, and applying Eq.~\ref{eq:varxi}
one would find
\begin{equation}
{\rm Var} (\xi) \approx \langle (2 \Delta_i - 0)^2 \rangle = 4 \, \langle \Delta_i^2 \rangle = 4 \, \frac{P_{\rm real}(0)}{L_{\rm box}^3} \frac{D^2(a_{\rm uni})}{D^2(1)}. \label{eq:ignorance}
\end{equation}
In the powerlaw models investigated in this section but certainly also in 
$\Lambda$CDM cosmologies the above result would be an order of magnitude {\it larger}
than any of the other sources of box-to-box variance just mentioned and many more 
simulations would need to be performed to measure the mean $\xi(r)$ with any 
kind of precision. This underscores the importance of using a ``better informed''
estimator for which $\xi_{{\rm uni},i}(r) = 0$ when the particles are uncorrelated. 
The cost of using an estimator that is ignorant of the overdensity is severe.

One may ask, then, what Eq.~\ref{eq:varxi} really means for $\xi$-sampled simulations
using a ``better informed'' estimator.
The answer is that the definition of the variance is not substantially changed. 
Eq.~\ref{eq:varxi} is the variance (or, equivalently, width of the distribution) 
of correlation function measurements at a particular separation, $r$, 
from finite volumes in a situation where the mean density
of the universe is perfectly known and the overdensities of each volume are also perfectly known. 
In other words, the only {\it unknown} is $\xi(r)$, which is what we are trying to measure.
This is exactly as it would be in a $P$-sampled ensemble of simulations where all overdensities
are perfectly known to be zero and the mean density of the universe is likewise perfectly known.
So although the task of measuring $\xi(r)$ is somewhat 
more complicated in $\xi$-sampled simulations (i.e. because of Eq.~\ref{eq:xiunii3} and 
the volumetric weighting), the measurement in principle is not qualitatively different 
from $P$-sampled ensembles. This being the case,
{\it the most important source of variance for both methods comes from 
the fact that we are trying to measure the mean correlation function of the universe from 
a finite number of randomly-realized density fields with known overdensities. }

\subsection{Expectations from Gaussian Statistics}
\label{sec:varbox}

Mindful that the correlation function is also the fourier transform of the power 
spectrum (Eq.~\ref{eq:ft}), the statement just made regarding the most
important source of variance can also be conveyed by pointing out
that finite volumes contain a finite number of fourier modes
that can be used to compute statistics like the correlation function. 
Since the number of modes in the simulation box for each $k$ value 
is straightforwardly determined this consideration can be used to 
estimate the variance of $\xi(r)$ using a linear theory approximation
for $P(k)$. Applying this reasoning one arrives at an estimate for the variance, 
\begin{equation}
\sigma_\xi^2 = \frac{1}{V \pi^2} \int_0^\infty {dk}k^2 \left(\frac{\sin kr}{kr} \right)^2 P(k)^2 \label{eq:xigaus}
\end{equation}
\citep{Cohn2006}. This is referred to as a ``gaussian'' estimate of the variance 
because in the approximation of gaussian random fields, wherein all higher order statistics (e.g. 3-point and 4-point functions) are assumed to be negligible, Eq.~\ref{eq:xigaus} 
perfectly models the variance of $\xi(r)$. For pure powerlaw models, $P(k) = Ak^n$, it can be shown using Eq.~\ref{eq:xigaus} that
\begin{equation}
\frac{\sigma_\xi}{\xi_{\rm{pow}}(r)} = \frac{A_n}{\pi} \frac{\sqrt{\Gamma(1+2n)\sin n \pi}}{4^{(n+1)/2}} \left( \frac{r}{L_{\rm{box}}} \right)^{3/2}  \label{eq:gausvar}
\end{equation}
where $A\equiv A_n r_0^{n+3}$, and $\Gamma(1+2n)$ is the usual gamma 
function. Notice that all of the $r_0$ dependence has canceled out with 
the division by $\xi_{\rm{pow}}(r) = (r_0 / r)^{n+3}$. Unfortunately, 
Eq.~\ref{eq:gausvar} is only convergent for the limited range of 
$-1.5 < n < -0.5$. 
\begin{figure*}
\centerline{\epsfig{file=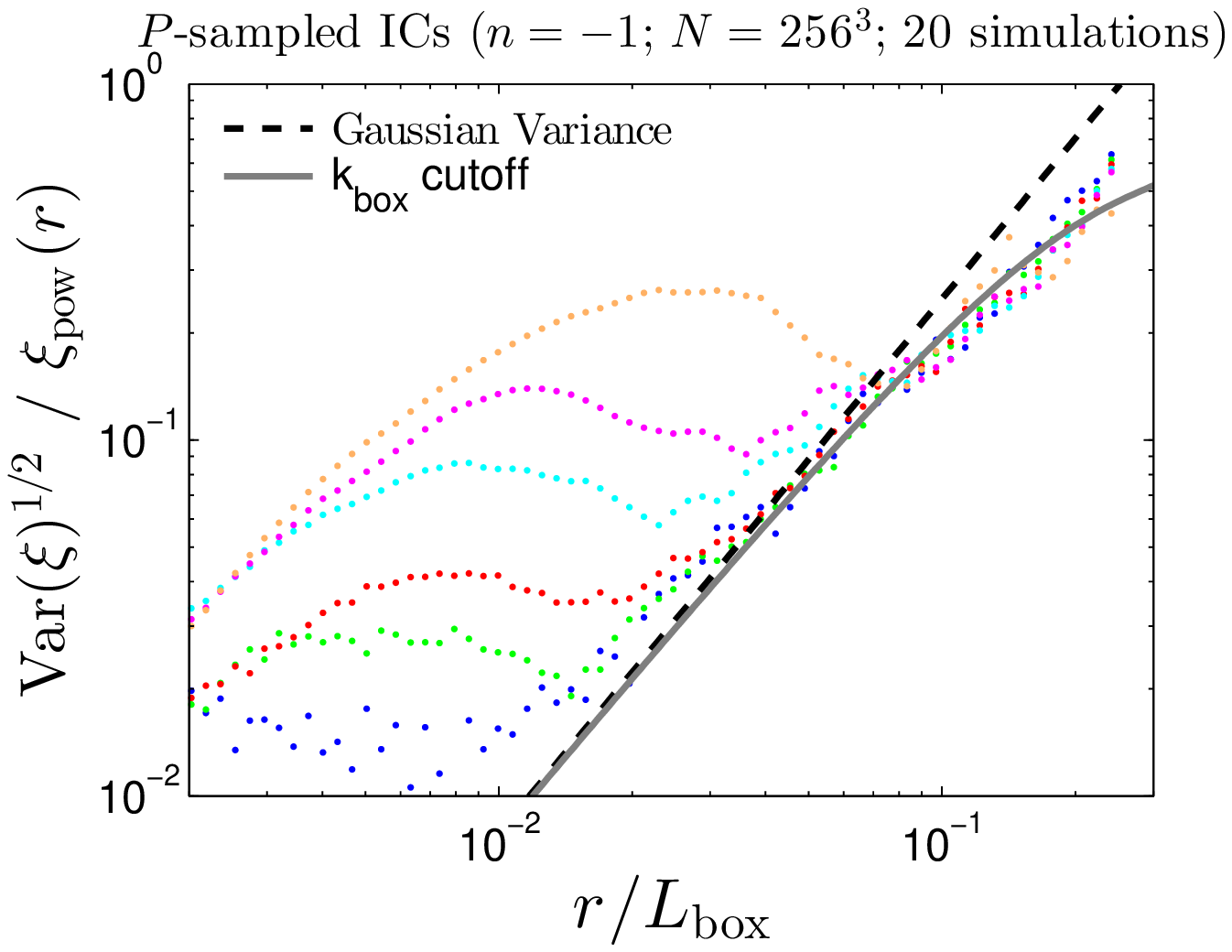, width = 3in}\epsfig{file=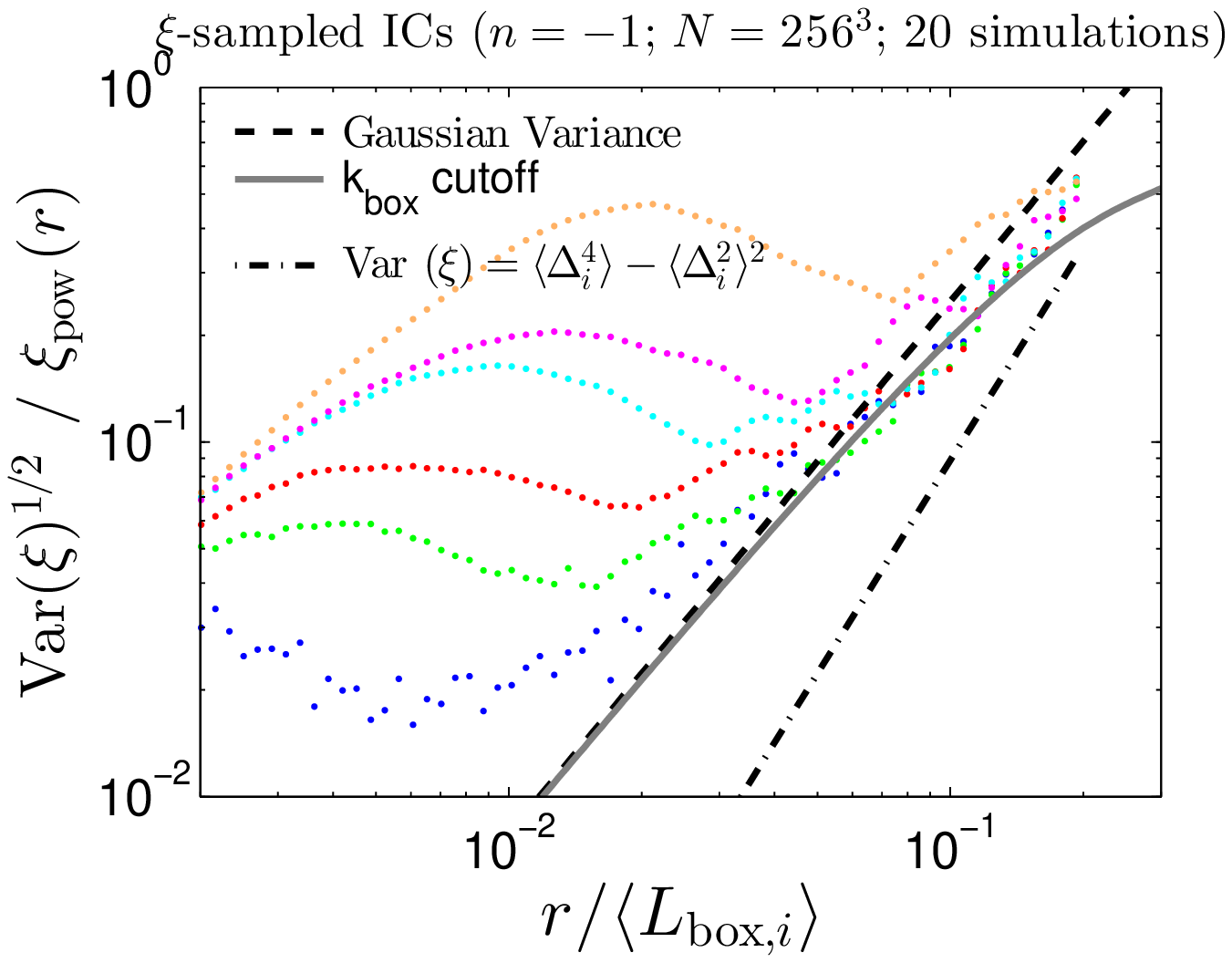, width=3in}}
\centerline{\epsfig{file=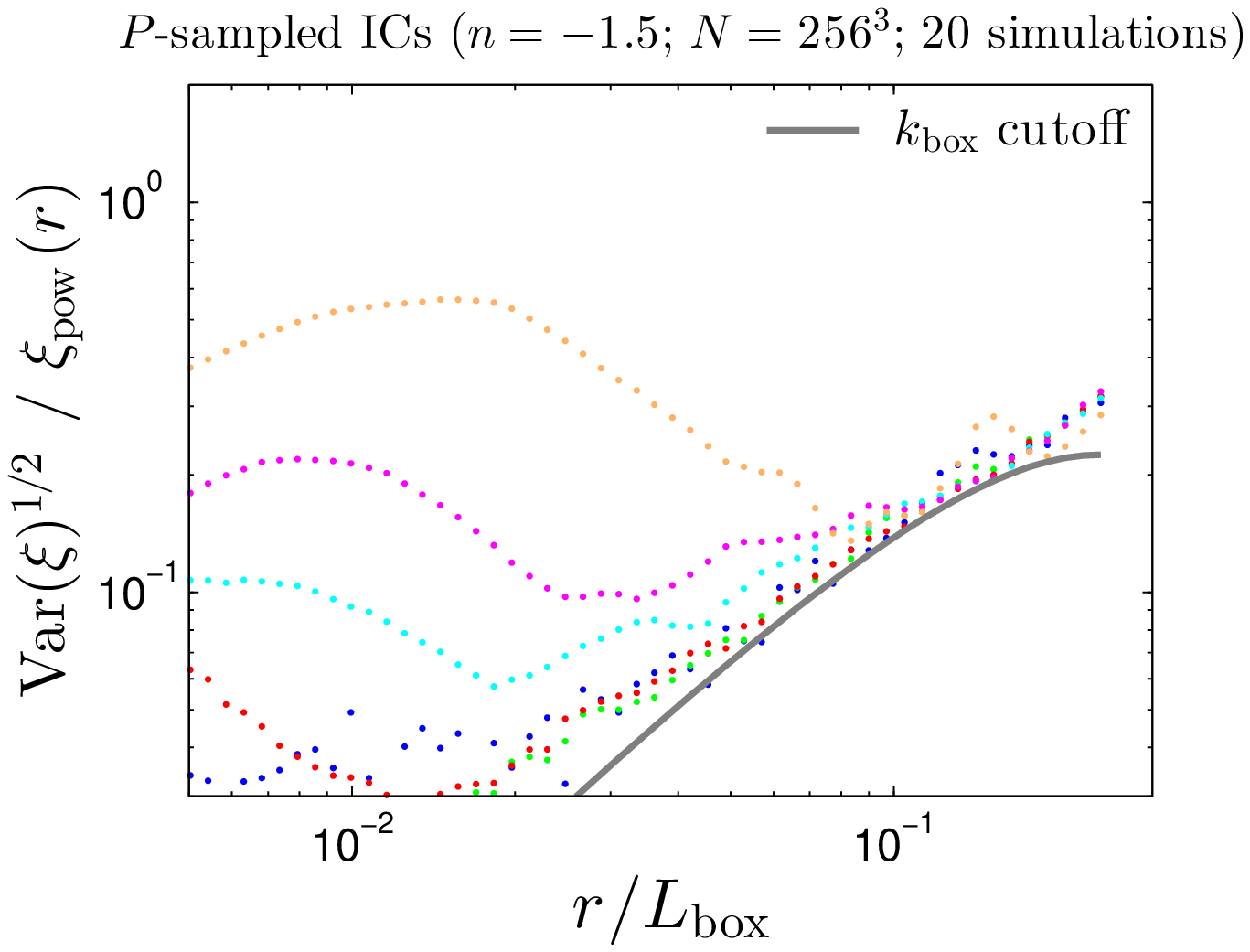, width = 3in}\epsfig{file=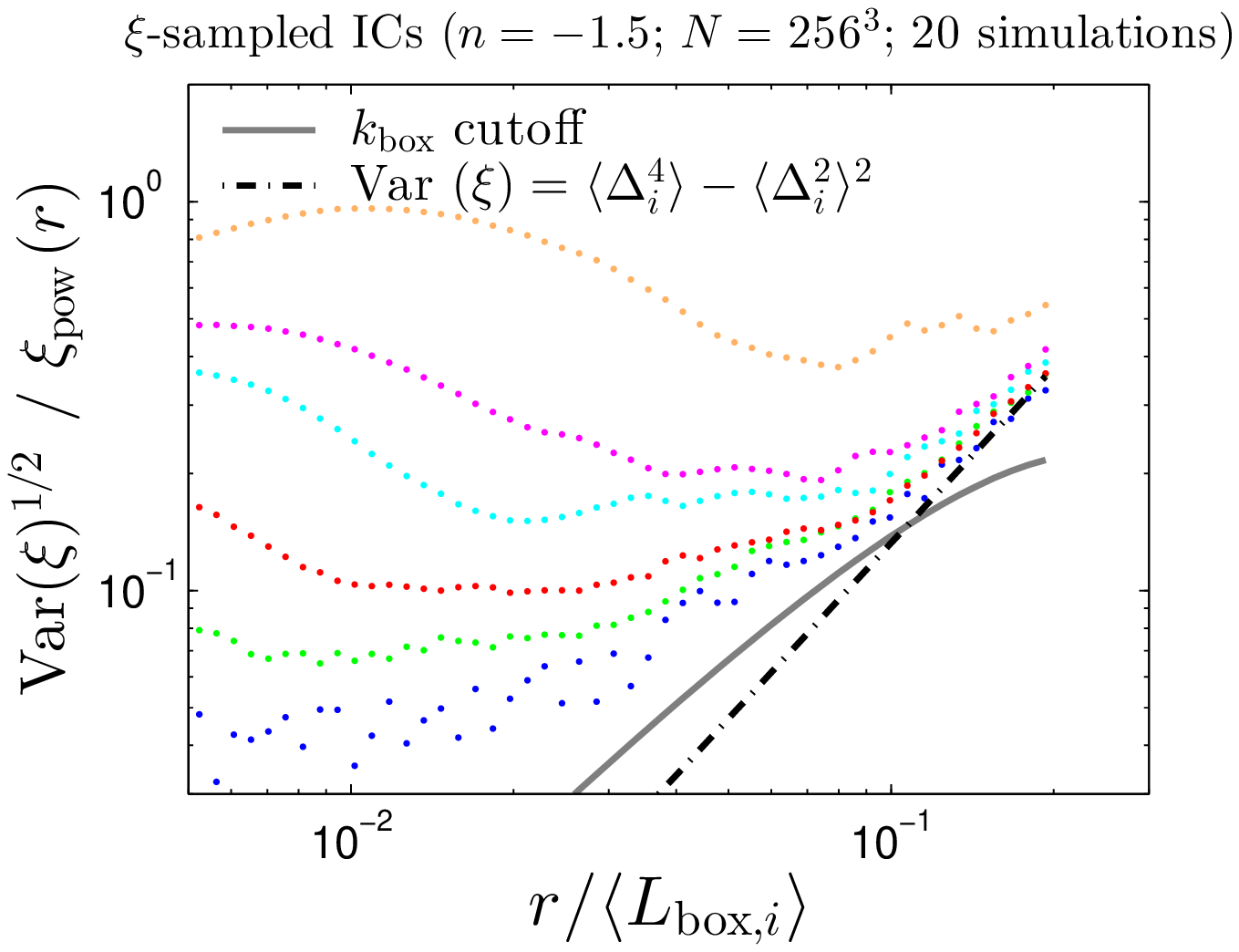, width=3in}}
\centerline{\epsfig{file=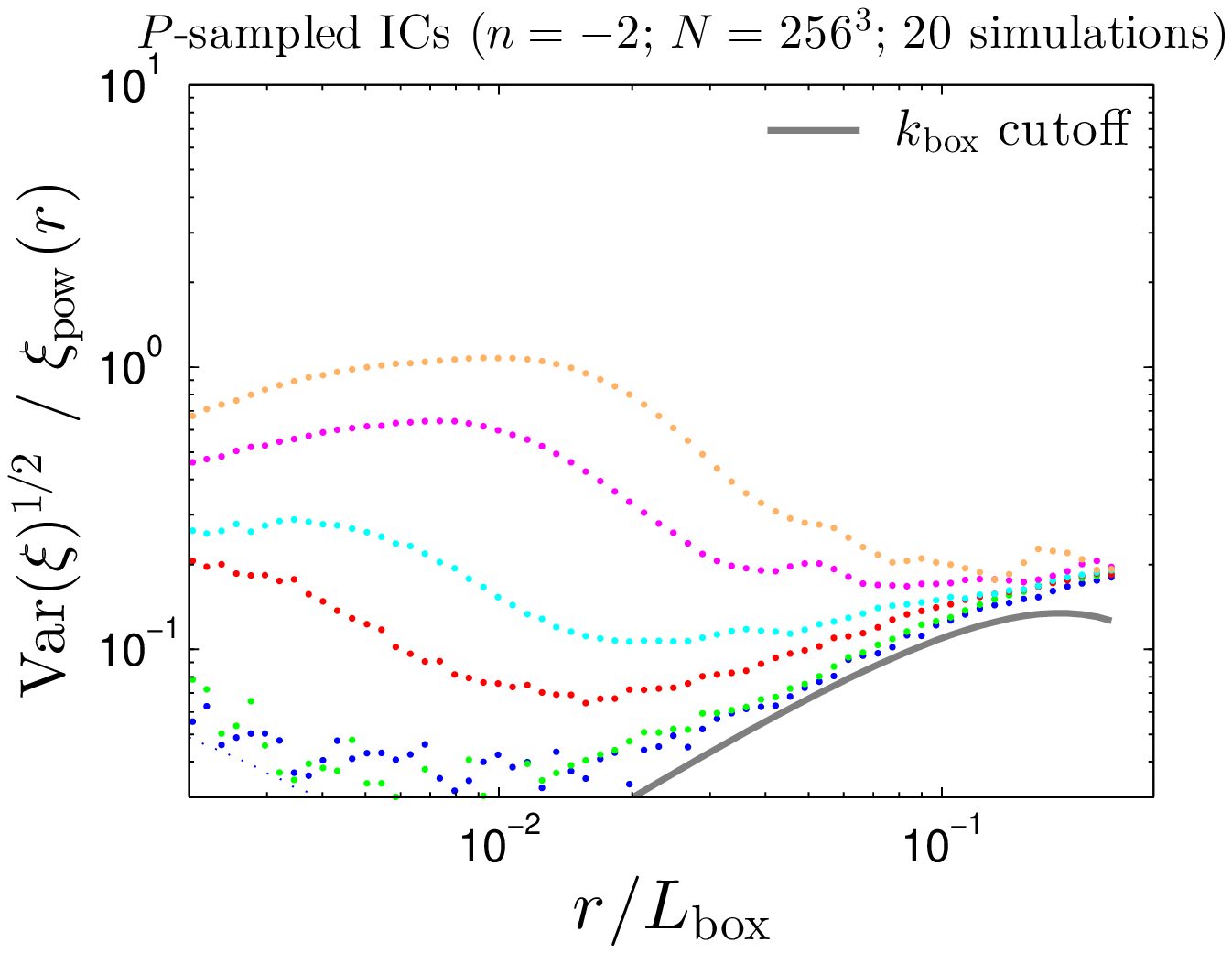, width = 3in}\epsfig{file=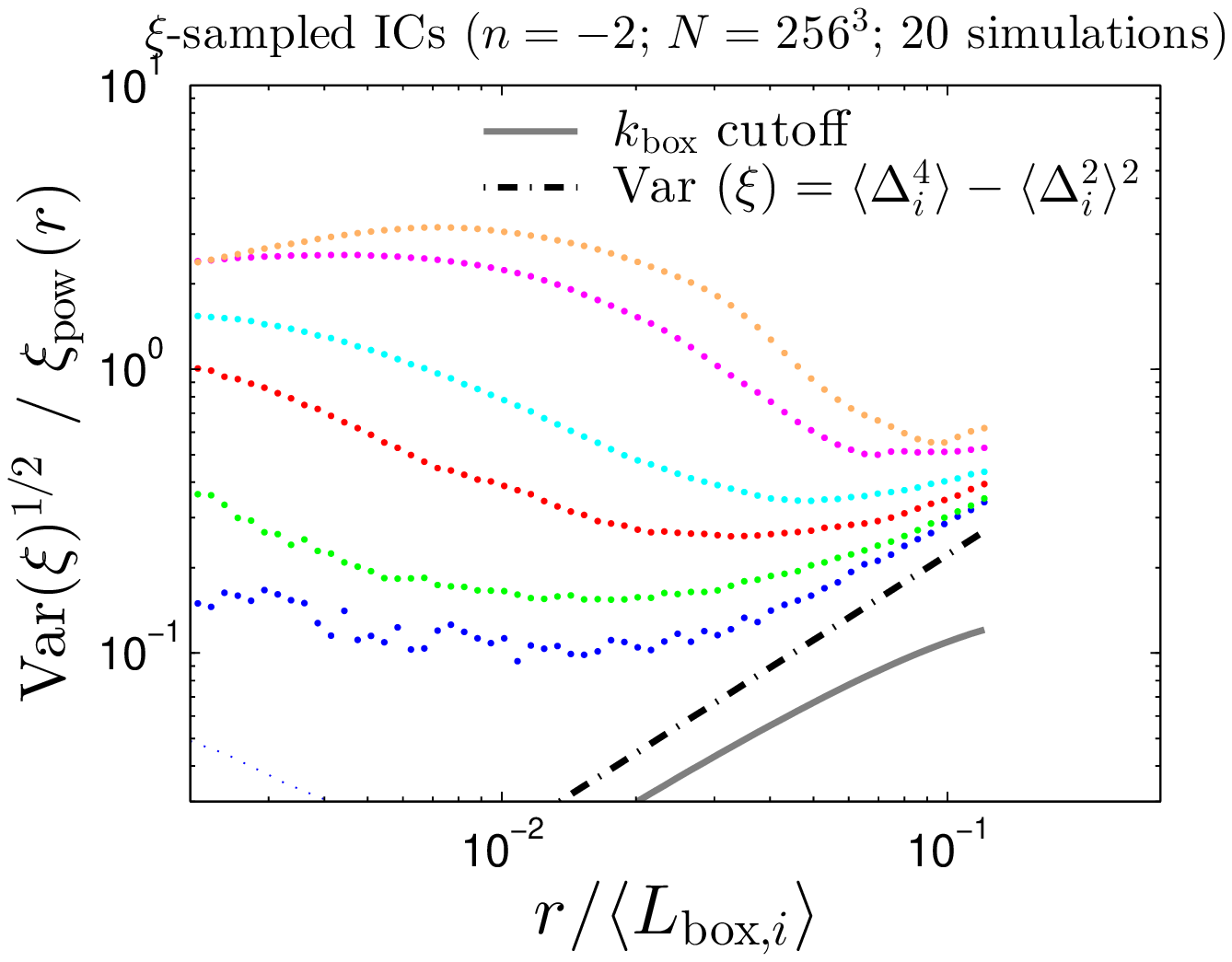, width=3in}}
\vspace{-0.1in}
\caption{ Measurements of the box-to-box variance of $\xi(r)$ from 
simulations (colored points in each panel, see Fig.~\ref{fig:xisamp} for legends) 
compared to expectations from gaussian statistics (Eq.~\ref{eq:xigaus} in 
dashed black lines, and Eq.~\ref{eq:xigaus} with a low-$k$ cutoff for the 
integral at $k_{\rm box} = 2 \pi / L_{\rm box}$ shown with solid 
gray lines). The $x$-axis shows the separation, $r$, relative to the scale of the 
simulation box. Also shown alongside measurements from $\xi$-sampled simulations
is an extra source of variance from Eq.~\ref{eq:extravar}. Note that 
the $\xi$-sampled measurements do not extend to separations as close 
to the box scale as the $P$-sampled measurements. In highly overdense boxes, 
this avoids measuring $\xi(r)$ for separations larger than $L_{{\rm box},i} /4$.
}\label{fig:xigaus}
\end{figure*}
Fig.~\ref{fig:xigaus} compares Eq.~\ref{eq:xigaus} with a low-$k$ cutoff at $k_{\rm box}$ 
(solid gray lines) to the box-to-box variance 
measured from the simulation ensembles in detail, showing the separation, $r$, 
relative to the scale of the box and normalizing the $y$-axis by $\xi_{\rm pow}(r)$ so
that the gaussian expectation of Eq.~\ref{eq:gausvar} is independent of epoch. 
For the convergent case of $n = -1$, Eq.~\ref{eq:gausvar} without a low-$k$ 
cutoff (dashed black lines) is also compared to the simulation data.
The $\xi$-sampled results are also compared to another source of 
variance (black dot-dashed lines, Eq.~\ref{eq:extravar}) that will be explained in the next section.

\subsection{Commentary on Figure~\ref{fig:xigaus}}
\label{sec:comments}

The $n = -1$ results in Fig.~\ref{fig:xigaus} are most instructive since Eq.~\ref{eq:xigaus}
is compared to the measured variance from simulations both with and without the low-$k$ cutoff.
In each panel in Fig.~\ref{fig:xigaus} the range $r / L_{\rm box} \gtrsim 1/10$ is most important 
for this comparison because on smaller scales and increasingly for later outputs the box-to-box 
variance from non-linear correlations, which are not accounted for in Eq.~\ref{eq:xigaus},
become important and greatly exceed the linear theory expectation of 
Eq.~\ref{eq:xigaus}\footnote{According to Hyper-Extended Perturbation Theory \citep[HEPT;][]{Scoccimarro_Frieman1996b} non-linear and higher order contributions to the box-to-box variance grow as 
\begin{equation}
\frac{\sigma_{\xi,\rm{hept}}}{\xi_{\rm{pow}}(r)} = \sqrt{ 4 (1 - 2Q_3 + Q_4) \, \bar{\xi}_L (R_{\rm{box}}) } \sim \left ( \frac{r_o}{L_{\rm box}} \right)^{(n+3)/2} \label{eq:hept}
\end{equation}
where $Q_3$ and $Q_4$ are constants from HEPT that depend on $n$. Notice that $r_o$ does {\it not} cancel out
as in Eq.~\ref{eq:gausvar}, so this source of variance grows larger as the simulation progresses.
More exactly, Eq.~\ref{eq:hept} predicts that $\sigma_{\xi} / \xi_{\rm pow}(r)$ on non-linear scales 
will increase in proportion to the linear growth function. This prediction was confirmed 
with a detailed comparison of Eq.~\ref{eq:hept} to the measurements from simulations in \cite{Orban2011} 
however HEPT was overall consistent with the simulations only at an order-of-magnitude level.
Note that the Journal version of \cite{Scoccimarro_Frieman1996b} contains a typo for $Q_4$. The arXiv version is correct or, c.f., \cite{Bernardeau_etal2002} (R. Scoccimarro private communication). }.
But for $r / L_{\rm box} \gtrsim 1/10$,  the $n = -1$ case measurements of the variance 
generically fall below Eq.~\ref{eq:xigaus} {\it without} the low-$k$ cutoff and 
are either consistent with or slightly above the expected variance 
from including the low-$k$ cutoff in Eq.~\ref{eq:xigaus}.
That both methods fall {\it below} the expectation of Eq.~\ref{eq:xigaus} without
the low-$k$ cutoff is a sensible result, especially for $P$-sampled
simulations because it is an explicit assumption of the method 
that $P(k) = 0$ for all $k$-modes from scales larger than the size of the simulation box. 
As a result, clustering power on these scales do not contribute to
the box-to-box variance of $\xi(r)$. 

The $\xi$-sampled $n = -1$ results falling below the expectation of Eq.~\ref{eq:xigaus} 
is also sensible for two reasons: 
(1) as argued earlier, the definition of the box-to-box variance is not 
substantially changed. And (2) despite including the fluctuations in the DC mode of the simulations,
one still expects the $\xi$-sampled method to under-represent clustering on scales larger
than the box. Notice, for example, that $P_{\rm real}(k)$ is totally insensitive to 
clustering power in $\xi(r)$ from $r > L_{\rm box} / 2$ (Eq.~\ref{eq:preal})
and it is $P_{\rm real}(k)$ that is used in the Zel'dovich formulism to generate 
the initial conditions.

Turning to the $n = -1.5$ and $-2$ results, clearly the measurements from
the $P$-sampled simulations for $n = -1.5$ and $-2$ also compare well to the expectation from
Eq.~\ref{eq:xigaus} with a cutoff at $k_{\rm box}$ as expected. 
However, the $\xi$-sampled results clearly exceed the expectation of Eq.~\ref{eq:xigaus}
with the low-$k$ cutoff. This stems from the fact that, compared to the $n = -1$
simulations, the fluctuations in the overdensity are larger for the $n = -1.5$ 
simulations and even larger for the $n = -2$ simulations (c.f. $P_{\rm real}(k \rightarrow 0)$ in Fig.~\ref{fig:pkcompare}).
But the key is that in $\xi$-sampled simulations the correction for 
the integral constraint bias arises naturally because $\xi_{{\rm uni},i}(r)~\rightarrow~\Delta_i^2$
on large scales where the particle distribution is approximately uncorrelated 
(Eq.~\ref{eq:Delta2}). When this is the case, the box-to-box variance (Eq.~\ref{eq:varxi}) yields
\begin{equation}
{\rm Var }(\xi) \approx \langle (\Delta_i^2 - \langle \Delta_i^2 \rangle )^2 \rangle = \langle \Delta_i^4 \rangle - \langle \Delta_i^2 \rangle^2. \label{eq:extravar}
\end{equation}
Since $\Delta_i = \frac{D(a_{\rm uni})}{D(1)}\Delta_{0,i}$ and $\Delta_{0,i}$ is a 
gaussian random variable, then $\langle \Delta_i^4 \rangle = 3 \langle \Delta_i^2 \rangle^2$ and
\begin{equation}
{\rm Var}(\xi) = 3 \langle \Delta_i^2 \rangle^2 - \langle \Delta_i^2 \rangle^2 = 2 \, \langle \Delta_i^2 \rangle^2 = 2 \, \left(\frac{P_{\rm real}(0)}{L_{\rm box}^3} \frac{D^2(a_{\rm uni})}{D^2 (1)} \right)^2 .
\end{equation}
Although the above expression is smaller than, e.g., Eq.~\ref{eq:ignorance} it can 
be as large or larger than Eq.~\ref{eq:xigaus}. So while the expectation of 
Eq.~\ref{eq:xigaus} with a low-$k$ cutoff compared well to the 
measurements from simulation in all the other panels, this is why the 
$\xi$-sampled $n = -1.5$ and $-2$ results in Fig.~\ref{fig:xigaus} so greatly exceed the expectation
from Eq.~\ref{eq:xigaus} with the low-$k$ cutoff even in the $r \gtrsim L_{\rm box} / 10$
region where non-linear effects are small. 

In $\xi$-sampled $\Lambda$CDM simulations, such as those in Sirko \cite{Sirko2005}, this issue would
likewise artificially increase the box-to-box variance and degrade the error on the 
mean $\xi(r)$. If the box size is small enough then $\langle \Delta_i^4 \rangle^{1/2}$ will
be comparable to $\sigma_\xi$ from Eq.~\ref{eq:xigaus}\footnote{Since $\sigma_\xi$ and 
$\langle \Delta_i^4 \rangle^{1/2}$ both become smaller for increasing $L_{\rm box}$ this is a non-trivial 
statement. From Eq.~\ref{eq:xigaus} or \ref{eq:gausvar}, $\sigma_\xi \propto L_{\rm box}^{-3/2}$ and, according to 
similar reasoning as employed in Appendix~\ref{ap:ximatter}, $P_{\rm real}(0) / L_{\rm box}^3 \sim L_{\rm box}^{-4}$. 
So it must always be true that for small enough $L_{\rm box}$ that $\langle \Delta_i^4 \rangle^{1/2} \gg \sigma_\xi$.} .
Indeed this seems to be the case in their Fig.~9 which presents $\xi$-sampled
simulations with $L_{\rm box} = 100~h^{-1}$~Mpc. The 1-sigma error on the mean $\xi(r)$
in that case is noticeably larger than the 1-sigma error on the mean from the 
$P$-sampled simulations. Sirko \cite{Sirko2005} does not comment on this interesting result.
The work here suggests that this is just a consequence of $\xi_{{\rm uni},i}(r) \approx \Delta_i^2$
on the scale of the simulation box and how large typical values of $\Delta_i^2$
can be for 100~$h^{-1}$~Mpc boxes (c.f. their Fig.~4). Perhaps in future investigations
a $\xi$-sampled estimator can be constructed to prevent this extra source of 
variance from contributing but without removing the  
compensation for the integral constraint bias (\S~\ref{sec:intbiasxi}). 
Viewed another way, this additional complication with $\xi$-sampled simulations
highlights the simplicity and economy of $P$-sampled simulations which
with only a small correction for the integral constraint (Eq.~\ref{eq:xibias}) yields
an unbiased estimate of the mean $\xi(r)$ and does so with a large-scale variance 
that corresponds very closely to the approximation of gaussian random fields as it should.

\section{Summary and Conclusions}
\label{sec:end}

This paper explores the predictions from both the conventional method of running 
ensembles of cosmological simulations and an alternative approach proposed 
by \cite{Pen1997} and implemented by \cite{Sirko2005}. The conventional method
is dubbed the $P$-sampled approach because it aims to maximize the correspondence
between the fourier space properties of the simulation and the fourier space
statistics of the assumed cosmological model whereas \cite{Pen1997} and \cite{Sirko2005}
outline a $\xi$-sampled approach which is built from focusing on real-space 
statistics. Unlike the conventional method, the real-space approach 
allows the DC mode to vary from box to box.  In an investigation comparing the 
$\xi$-sampled and $P$-sampled methods for the growth and evolution of
the matter-matter two-point correlation function the following conclusions
were drawn: \\

(1) Both $P$-sampled and $\xi$-sampled simulations give rise to the expected
self-similar behavior from powerlaw initial conditions 
(specifically $n = -1, -1.5$ \& $-2$). In the absence of exact solutions
for the non-linear growth of structure these tests robustly evaluate
the accuracy of the simulation method without assuming one or the 
other approach is correct \cite{Efstathiou_etal1985}.

(2) $\xi$-sampled simulations of BAO-inspired ``powerlaw times a bump'' 
models \cite{Orban_Weinberg2011} yielded consistent results with 
earlier, higher-resolution $P$-sampled simulations for the broadening and shift of
the BAO feature, even into the deeply non-linear regime ($\sigma_8 \gtrsim 1$).
A small but statistically significant discrepancy with the amplitude of the 
bump at early times can be attributed to a resolution effect.

(3) The earlier claim in Sirko 2005 \cite{Sirko2005} that the $\xi$-sampled
method performs better than the $P$-sampled method in modeling
the mean $\xi(r)$ in $\Lambda$CDM simulations for separations approaching
the box scale is incorrect because of an overlooked integral-constraint
correction to the $P$-sampled results presented there.
Appendix~\ref{ap:ximatter} derives a simple, independent-of-epoch analytic 
formula for estimating the importance of the integral-constraint bias in $\Lambda$CDM
simulations given the box size.

(4) In Fig.~9 of Sirko \cite{Sirko2005}, the error on the mean $\xi(r)$
for $\Lambda$CDM simulations was noticeably larger in $\xi$-sampled
simulations compared to $P$-sampled simulations. Sirko \cite{Sirko2005} 
did not comment on this interesting result. Investigations 
with powerlaw initial conditions show
that this larger variance comes from the behavior of the 
estimator on large scales where the particles are approximately
uncorrelated. Otherwise, the results both methods compare well to the ``gaussian''
expectation of the variance in Eq.~\ref{eq:varxi} because 
both estimators implicitly have perfect knowledge of the overdensities.

(5) A previously un-noticed constraint on initial conditions for $\xi$-sampled 
simulations requires that $n \geq -2 \,$ or, more generally, $n_{\rm eff} \geq -2$,
in order to keep the initial power spectrum positive (or equal to zero).
For $\Lambda$CDM simulations this forces $L_{\rm box} \geq 2.5 h^{-1}$ Mpc. \\

Now that the ensemble-averaged predictions for the correlation function using  
the $\xi$-sampled method have been explored and validated in some depth, future investigations
with the $\xi$-sampled method would do well to explore the ensemble-averaged 
predictions for halo clustering, halo mass functions and the power spectrum.
Indeed, there may be a statistic of interest for which including the 
fluctuations in the DC mode or some other aspect of the $\xi$-sampled
method is of particular importance \cite{Pen1997,Gnedin_etal2011}.

\section*{Acknowledgements}

The author thanks the Ohio State University Center for Cosmology and
AstroParticle Physics for its support, and David Weinberg for 
guidance. Thanks also goes to Jeremy Tinker for insightful conversations, Ed Sirko for helpful
correspondence and an anonymous referee who clarified some 
conceptual issues.  A special thanks to Stelios Kazantzidis (CCAPP) and the OSU 
astronomy department for making available compute nodes for this project, 
as well as the Ohio Supercomputer Center which was also a valuable resource. 
This project has been supported by NSF grant AST-1009505 and  AST-0707985.

\appendix 

\section{A Simple Expression for the Integral Constraint Bias in $\Lambda$CDM Simulations}
\label{ap:ximatter}

A simple derivation can be used to estimate the bias 
introduced by the integral constraint for large boxes assuming a $\Lambda$CDM initial 
power spectrum. In this case,
\begin{eqnarray}
\xi_{\rm bias}  = &\displaystyle - \frac{3}{4 \pi R_{\rm{box}}^3} \int_0^{R_{\rm{box}}  = L_{\rm box} / 1.61} 4 \pi r^2 \xi_{\Lambda CDM}(r) {dr} \nonumber \\
 = & \displaystyle - \frac{3}{ 4 \pi R_{\rm{box}}^3} \left[ \int_0^{\infty} 4 \pi r^2 \xi_{\Lambda CDM}(r) {dr}  -\int_{R_{\rm{box}}}^{\infty} 4 \pi r^2 \xi_{\Lambda CDM}(r) {dr} \right].
\end{eqnarray}
The integral over infinity is equivalent to $P(k \rightarrow 0)$ which goes to zero because $P(k) \sim k$ on large 
scales. The other term within the brackets can be approximated analytically since on scales larger than  $r \sim 250 h^{-1}$ Mpc, $\xi_{\Lambda CDM} \approx \xi_* (r_*/r)^{4}$ where $r_*$ is a constant and the amplitude, $\xi_*$, 
is negative. It can be easily shown that
\begin{eqnarray}
\xi_{\rm bias} \approx  3 \, \xi_* \left( \frac{r_*}{R_{\rm{box}}} \right)^4 = 20.16 \, \xi_*  \left( \frac{r_*}{L_{\rm{box}}} \right)^4.
\end{eqnarray}
Applying this result to estimate the fractional bias in the {\it amplitude} of the BAO feature yields 
\begin{eqnarray}
\frac{\xi (r_{\rm{bao}}) - \hat{\xi}(r_{\rm{bao}})} {\xi (r_{\rm{bao}})} = \frac{-\xi_{\rm bias}}{\xi (r_{\rm{bao}})} \approx 0.54 \% \left( \frac{ 1 h^{-1} \rm{Gpc}}{L_{\rm{box}}} \right)^4 \label{eq:lcdmbias}
\end{eqnarray}
where $\hat{\xi}(r_{\rm{bao}})$ is the uncorrected correlation function and I have assumed $(-\xi_*) / \xi (r_{\rm{bao}}) \approx$ 2.71e-4 and $r_* \approx 1 h^{-1}$ Gpc using 
CAMB \cite{camb_reference} and parameters from WMAP7 \cite{wmap7_reference}. 
Formally, because of a cancellation of the square of the linear theory 
growth function in the ratio $(-\xi_0) / \xi (r_{\rm{bao}})$, Eq.~\ref{eq:lcdmbias}
 is independent of redshift and, if left unaccounted for, this measurement bias will propagate 
to change inferences regarding the broadening and shift of the BAO feature in the correlation 
function as well regardless of epoch. In more detail, redshift-dependent contributions 
arising from higher-order correlations can also bias the correlation function \cite{Bernardeau_etal2002},
 so in practice Eq.~\ref{eq:lcdmbias} can be thought of as a {\it lower} bound. 

\bibliography{ms}
\bibliographystyle{apsrev} 

\end{document}